\documentclass[10pt, prb, aps, twocolumn, showpacs, citeautoscript, floatfix, reprint, amsmath, amssymb, notitlepage, superscriptaddress]{revtex4-1}

\usepackage{graphicx}
\usepackage{stmaryrd}
\usepackage{rotating}
\usepackage{amsmath}
\usepackage{amsfonts}
\usepackage{amssymb}
\usepackage{wasysym}
\usepackage[countmax]{subfloat}
\usepackage{dcolumn} 
\usepackage{bm} 
\usepackage{color}

\usepackage{float}
\usepackage{latexsym}

\begin{document}

\title{Quantum geometrical properties of topological materials}






\author{Wei Chen}
\email{wchen@puc-rio.br}
\affiliation{Department of Physics, PUC-Rio, 22451-900 Rio de Janeiro, Brazil}

\date{\rm\today}

\begin{abstract}

The momentum space of topological insulators and topological superconductors is equipped with a quantum metric defined from the overlap of neighboring valence band states or quasihole states. We investigate the quantum geometrical properties of these materials within the framework of Dirac models and differential geometry. The Ricci scalar is found to be a constant throughout the whole Brillouin zone, and the vacuum Einstein equation is satisfied in 3D with a finite cosmological constant. For linear Dirac models, several geometrical properties are found to be independent of the band gap, including the straight line geodesic, constant volume of the curved momentum space, exponential decay form of the nonlocal topological marker, and unity Euler characteristic in 2D, indicating the peculiar yet universal quantum geometrical properties of these models.


\end{abstract}

\maketitle

\section{Introduction}

From a mathematical perspective, topology is one of the characteristic geometrical properties of a compact manifold. Depending on the dimension of the manifold, the topology is formulated in different ways. For instance, the Gauss-Bonnet theorem states that the integration of Gaussian curvature over a two-dimensional (2D) compact Euclidean manifold gives the number of handles it contains, a theorem that ubiquitously applies to the surface of almost all three-dimensional (3D) objects in our daily life. Following this mathematical perspective, it is then natural to speculate that the topological order in solids may also be viewed as a differential geometrical property of a certain manifold\cite{Hasan10,Qi11}. In particular, the topological order in topological insulators (TIs) and topological superconductors (TSCs) are generally calculated from the momentum space integration of a certain curvature function, such as Berry connection\cite{Zak89} or Berry curvature\cite{Thouless82,Berry84} of the valence band states or quasihole states, over the Brillouin zone (BZ) which is a compact manifold of $T^{D}$ torus. This observation has triggered a fair amount of interest on the geometrical aspects of the momentum space Euclidean manifold of TIs and TSCs, especially on the search of a proper definition of geometry that is relevant to the topological order.

A scenario that emerges to be the proper geometrical description relevant to the topological order is the quantum geometry of the filled states. This notion of quantum geometry arises from the quantum metric of the fully antisymmetric filled band state that respects the Fermi statistics of electrons or Bogoliubov quasiparticles, where one uses the overlap of two such states at slightly different momenta to define a metric\cite{Provost80}, which can also be regarded as a fidelity susceptibility treating momentum as a parameter ${\bf k}$\cite{You07,Zanardi07,Gu08,Yang08,Albuquerque10,Gu10,Carollo20}. Within the framework of Dirac models that describe prototype TIs and TSCs, the quantum metric has been shown to be related to topological order in various ways. For example, early investigations show that the modulus of the Berry connection in certain 1D systems and that of the Berry curvature in certain 2D systems are directly proportional to the square root of the determinant of the quantum metric\cite{Ma13,Ma14,Yang15,Piechon16,Panahiyan20_fidelity}. Recently, based a universal topological invariant applicable to TIs and TSCs in any dimension and symmetry class\cite{Schnyder08,Ryu10,Kitaev09,Chiu16,vonGersdorff21_unification}, it is uncovered that the modulus of the curvature function that momentum-integrates to the invariant is always proportional to the square root of the determinant\cite{vonGersdorff21_metric_curvature}. Furthermore, since the square root of the determinant is also the volume form, this naturally leads to the conclusion that the universal topological invariant is bounded by the volume of the curved momentum space\cite{Mera22}.

Motivated by these pioneering works that demonstrate the relation between quantum geometry and topological order, in the present work we use an analytical expression for the quantum metric to investigate a number of differential geometrical properties of Dirac models, including Christoffel symbol, Riemann tensor, Ricci scalar, Einstein equation, cosmological constant, and geodesic. In contrast to other types of systems\cite{Kolodrubetz13,Kolodrubetz17,Smith22}, we find that topological materials have a number of universal geometrical properties that are independent of the parameters of the Dirac Hamiltonian: Within linear Dirac models, the geodesic, understood as the trajectory in the momentum space along which the valence band/quasihole state as a unit vector in the Hilbert space rotates the least, is always a straight line. The Ricci scalar is found to be always constant and only determined by the dimension of the Dirac Hamiltonian, and the Euler characteristics of 2D topological materials obtained from the momentum integration of volume form times Ricci scalar is always unity. Besides, the vacuum Einstein equation is satisfied in 3D with a finite cosmological constant. Finally, the Fourier transform of the curvature function, equivalently a nonlocal topological marker that serves as a proper correlation function to detect topological phase transitions (TPTs)\cite{Chen23_universal_marker}, satisfies an exponential decay form. We further use lattice Dirac models to numerically investigate these geometrical properties in realistic materials, the investigation of which reveals a rather peculiar dependence of quantum metric on the material parameters.


\section{Quantum geometry of Dirac models}

\subsection{Quantum geometry}

Our goal is to study the quantum geometry of generic topological materials in any topologically nontrivial symmetry classes in $D$-dimension described by the Dirac Hamiltonian\cite{Schnyder08,Ryu10,Chiu16}
\begin{eqnarray}
H({\bf k})=\sum_{i=0}^{D}d_{i}\Gamma_{i},
\label{generic_Dirac_models}
\end{eqnarray}
where $\Gamma_{i}$'s are the $N\times N$ Dirac matrices, and we denote $d=\left(\sum_{i=0}^{D}d_{i}^{2}\right)^{1/2}$ as the modulus of the ${\bf d(k)}=(d_{0},d_{1}...d_{D})$ vector that parametrizes the momentum dependence of the Hamiltonian. A universal topological invariant has been proposed to describe the topological order, which is calculated from momentum-integration of the cyclic derivative of the ${\bf d}$-vector\cite{vonGersdorff21_unification}
\begin{eqnarray}
{\rm deg}[{\bf n}]&=&\frac{1}{V_{D}}\int d^{D}{\bf k}\,\varepsilon_{i_{0}...i_{D}}\frac{1}{d^{D+1}}d_{i_{0}}\partial_{1}d_{i_{1}}...\partial_{D}d_{i_{D}}
\nonumber \\
&\equiv&\frac{1}{V_{D}}\int d^{D}{\bf k}\,J_{\bf k},
\label{wrapping_number_formula}
\end{eqnarray}
where $\varepsilon_{i_{0}...i_{D}}$ is the Levi-Civita symbol. The invariant ${\rm deg}[{\bf n}]$ in Eq.~(\ref{wrapping_number_formula}) has been referred to as the wrapping number or degree of the map, which counts the number of times the $T^{D}$ BZ wraps around the unit sphere $S^{D}$ that the unit vector ${\bf n}={\bf d}/d$ forms, and the integrand $J_{\bf k}$ is the Jacobian of the map\cite{vonGersdorff21_unification}. Here $V_{D}=2\pi^{(D+1)/2}/\Gamma(\frac{D+1}{2})$ is the volume of the $D$-sphere of unit radius that takes the values $\left\{V_{1},V_{2},V_{3}...\right\}=\left\{2\pi,4\pi,2\pi^{2}...\right\}$, and $\partial_{\mu}\equiv\partial/\partial k^{\mu}$ is the derivative over contravariant momentum $k^{\mu}$. In addition, the Fourier transform of the Jacobian is also of particular interest
\begin{eqnarray}
\tilde{F}_{D}({\bf R})=\frac{1}{V_{D}}\int d^{D}{\bf k}\,J_{\bf k}e^{i{\bf k\cdot R}}={\cal C}({\bf r+R,r}),
\end{eqnarray}
since it gives a nonlocal topological marker\cite{Chen23_universal_marker} ${\cal C}({\bf r+R,r})$ that can also be interpreted as a Wannier state correlation function\cite{Chen17,Chen19_universality,Chen19_AMS_review,Molignini23_Chern_marker,Chen23_spin_Chern_marker}. We shall see below that this nonlocal marker decays in real space with a decay length the diverges at the TPTs, thereby serving as a faithful correlator to detect the critical behavior near the TPT.

All the TIs and TSCs under question contain $N_{-}$ valence band states or quasihole states denoted by $|n^{\bf k}\rangle$, and $N_{+}$ conduction band states or quasiparticle states denoted by $|m^{\bf k}\rangle$, where $N_{-}+N_{+}=N$. To respect the Fermi statistics, we consider the fully antisymmetric valence band or quasihole Bloch state at momentum ${\bf k}$ that is filled with electrons or quasiparticles, described by 
\begin{eqnarray}
|u^{\rm val}({\bf k})\rangle=\frac{1}{\sqrt{N_{-}!}}\epsilon^{n_{1}n_{2}...n_{N-}}|n_{1}^{\bf k}\rangle|n_{2}^{\bf k}\rangle...|n_{N_{-}}^{\bf k}\rangle,\;\;\;
\label{psi_val}
\end{eqnarray}
where we aim to investigate the quantum metric of this state defined from\cite{Provost80} (repeated Greek indices $\left\{\mu,\nu\right\}$ are summed)
\begin{eqnarray}
|\langle u^{\rm val}({\bf k})|u^{\rm val}({\bf k+\delta k})\rangle|=1-\frac{1}{2}g_{\mu\nu}({\bf k})\delta k^{\mu}\delta k^{\nu},
\label{uval_gmunu}
\end{eqnarray}
Note that we have used the contravariant momentum $k^{\mu}$ as the momentum in the usual sense, which is distinct from the covariant momentum $k_{\mu}$ obtained from the former via 
\begin{eqnarray}
k_{\mu}=g_{\mu\nu}k^{\nu},\;\;\;
k^{\mu}=g^{\mu\nu}k_{\nu},\;\;\;
g_{\mu\nu}g^{\nu\rho}=\delta_{\mu}^{\rho},
\end{eqnarray}
and hence in practical calculations the $g^{\mu\nu}$ is the inverse matrix of $g_{\mu\nu}$. The quantum metric has the expression\cite{vonGersdorff21_metric_curvature} 
\begin{eqnarray}
&&g_{\mu\nu}({\bf k})=\frac{1}{2}\langle \partial_{\mu}u^{\rm val}|\partial_{\nu}u^{\rm val}\rangle+\frac{1}{2}\langle \partial_{\nu}u^{\rm val}|\partial_{\mu}u^{\rm val}\rangle
\nonumber \\
&&-\langle \partial_{\mu}u^{\rm val}|u^{\rm val}\rangle \langle u^{\rm val}|\partial_{\nu}u^{\rm val}\rangle
\nonumber \\
&&=\frac{1}{2}\sum_{nm}\left[\langle \partial_{\mu}n|m\rangle\langle m|\partial_{\nu}n\rangle+\langle \partial_{\nu}n|m\rangle\langle m|\partial_{\mu}n\rangle\right].
\label{gmunu_T0}
\end{eqnarray}
Furthermore, for Dirac models of the form of Eq.~(\ref{generic_Dirac_models}), the quantum metric is\cite{vonGersdorff21_metric_curvature} 
\begin{eqnarray}
g_{\mu\nu}({\bf k})=\frac{N}{8d^{2}}\left\{\sum_{i=0}^{D}\partial_{\mu}d_{i}\partial_{\nu}d_{i}
-\partial_{\mu}d\partial_{\nu}d\right\}.
\label{quantum_metric_general_formula}
\end{eqnarray}
which is valid for any coordinate and any parametrization ${\bf d}$. Furthermore, a relation between the determinant of the quantum metric and the modulus of the Jacobian in Eq.~(\ref{wrapping_number_formula}) has been discovered
\begin{eqnarray}
|J_{\bf k}|=\left(\frac{8}{N}\right)^{D/2}\sqrt{\det g_{\mu\nu}},
\label{metric_curvature_correspondence}
\end{eqnarray}
which has been referred to as the metric-curvature correspondence that holds for any parametrization ${\bf d}({\bf k})$ of the Hamiltonian\cite{vonGersdorff21_metric_curvature}.

Our interest is to use the analytical formula in Eq.~(\ref{quantum_metric_general_formula}) to derive the differential geometrical quantities that describe the Euclidean manifold of momentum space, such as the Christoffel symbol $\Gamma_{\mu\nu}^{\lambda}$, Riemann tensor $R^{\rho}_{\;\sigma\mu\nu}$, Ricci Tensor $R_{\mu\nu}$, Ricci scalar $R$, and Einstein tensor $G_{\mu\nu}$ defined by\cite{Carroll03}
\begin{eqnarray}
&&\Gamma_{\mu\nu}^{\lambda}=\frac{1}{2}g^{\lambda\sigma}(\partial_{\mu}g_{\nu\sigma}
+\partial_{\nu}g_{\sigma\mu}-\partial_{\sigma}g_{\mu\nu}),
\nonumber \\
&&R^{\rho}_{\;\sigma\mu\nu}=\partial_{\mu}\Gamma_{\nu\sigma}^{\rho}-\partial_{\nu}\Gamma_{\mu\sigma}^{\rho}
+\Gamma_{\mu\lambda}^{\rho}\Gamma_{\nu\sigma}^{\lambda}-\Gamma_{\nu\lambda}^{\rho}\Gamma_{\mu\sigma}^{\lambda},
\nonumber \\
&&R_{\mu\nu}=R^{\lambda}_{\;\mu\lambda\nu},\;\;\;
R=g^{\mu\nu}R_{\nu\mu},
\nonumber \\
&&G_{\mu\nu}=R_{\mu\nu}-\frac{1}{2}Rg_{\mu\nu},
\label{Christoffel_Riemann_Ricci}
\end{eqnarray} 
from which we can study the geodesic equation in the momentum space parametrized by a variable $t$
\begin{eqnarray}
\frac{d^{2}k^{\mu}}{dt^{2}}+\Gamma_{\rho\sigma}^{\mu}\frac{dk^{\rho}}{dt}\frac{dk^{\sigma}}{dt}=0.
\end{eqnarray} 
Physically, because the metric defined in Eq.~(\ref{uval_gmunu}) measures how the state $|u^{\rm val}({\bf k})\rangle$ as a unit vector in the Hilbert space rotates as one moves from ${\bf k}$ to ${\bf k+\delta k}$, the geodesic can be understood as the trajectory in momentum space along which the $|u^{\rm val}({\bf k})\rangle$ rotates the least in the Hilbert space. Note that Eq.~(\ref{quantum_metric_general_formula}) implies that the metric is symmetric in the two indices, leading to a torsion-free Christoffel symbol
\begin{eqnarray}
g_{\mu\nu}=g_{\nu\mu},\;\;\;\Gamma_{\mu\nu}^{\lambda}=\Gamma_{\nu\mu}^{\lambda},
\end{eqnarray}
which simplifies the calculation of Eq.~(\ref{Christoffel_Riemann_Ricci}).

We remark that in semiconductors and insulators, the quantum metric is experimentally measurable from momentum-resolved optical absorption rate, which is feasible by pump-probe experiments combined with time- and angle-resolved photoemission spectroscopy (trARPES)\cite{vonGersdorff21_metric_curvature}. Thus shall the experiment be able to map out the momentum profile of all the components of the quantum metric $g_{\mu\nu}$ in the whole BZ, it is in principle possible to calculate the derivatives of $g_{\mu\nu}$ over momentum and obtain all the geometrical quantities in Eq.~(\ref{Christoffel_Riemann_Ricci}). In addition, the quantum metric integrated over momentum ${\cal G}_{\mu\nu}=\int\frac{d^{D}{\bf k}}{(2\pi)^{D}}\,g_{\mu\nu}$, of which we call the fidelity number\cite{deSousa23_fidelity_marker}, is particularly simple to measure. The trace ${\rm Tr}\,{\cal G}_{\mu\nu}$ of the fidelity number is of particular interest, since it gives the spread of valence band Wannier function that has been heavily discussed in the density functional theory\cite{Souza00,Marzari97,Marzari12}. It is recently revealed that this trace and hence the spread can be simply measured from the opacity of 2D semiconductors and the dielectric function in 3D semiconductors\cite{CardenasCastillo24_spread_Wannier}. Remarkably, the dielectric function of a prototype 3D TI, namely Bi$_{2}$Te$_{3}$, has been measured decades ago\cite{Greenaway65}, from which ${\rm Tr}\,{\cal G}_{\mu\nu}=0.747\hbar/\AA$ has been extracted without any fitting parameter.

Our aim in the following sections is to systematically study the topological and quantum geometrical properties of linear and lattice Dirac models from 1D to 3D. For this purpose, we consider the linear Dirac models in $D$-dimension parametrized by
\begin{eqnarray}
d_{0}=M,\;\;\;d_{i}=vk^{i},\;\;\;{\rm for}\;i=1\sim D,
\label{generic_linear_Dirac}
\end{eqnarray}
where $M$ is the mass term and $v$ is the Fermi velocity, and calculate the geometrical quantities in Eq.~(\ref{Christoffel_Riemann_Ricci}) conveniently in the polar and spherical coordinates. For lattice Dirac models, we consider\cite{Chen20_absence_edge_current}
\begin{eqnarray}
&&d_{0}=M+2DB-\sum_{i=1}^{D}2B\cos k^{i}a,\;\;\;
\nonumber \\
&&d_{i}=A\sin k^{i}a,\;\;\;{\rm for}\;i=1\sim D,
\label{generic_lattice_Dirac}
\end{eqnarray}
where $\left\{A,B\right\}$ are kinetic parameters, whose geometrical properties can be numerically calculated in the Cartesian coordinates.

\section{1D topological materials}

We now use the generic 1D topological materials in any of the five topologically nontrivial symmetry classes AIII, BDI, D, DIII, and CII as examples, which are described by the Dirac models in Eq.~(\ref{generic_Dirac_models}) with $D=1$.

\subsection{Quantum geometry of 1D linear Dirac model}

We first examine the 1D linear Dirac model parametrized by
\begin{eqnarray}
d_{0}=M,\;\;\;d_{1}=vk,
\end{eqnarray}
whose quantum metric according to Eq.~(\ref{quantum_metric_general_formula}) and Jacobian according to Eq.~(\ref{wrapping_number_formula}) are 
\begin{eqnarray}
g_{kk}=\frac{NM^{2}v^{2}}{8(M^{2}+v^{2}k^{2})^{2}},\;\;\;
J_{k}=\frac{Mv}{(M^{2}+v^{2}k^{2})},
\end{eqnarray}
satisfying the metric-curvature correspondence in Eq.~(\ref{metric_curvature_correspondence}). The total volume of momentum space
\begin{eqnarray}
V_{1D}^{\rm lin}=\int_{-\infty}^{\infty} dk\sqrt{g_{kk}}=\pi\sqrt{\frac{N}{8}},
\end{eqnarray}
is independent of the band gap $M$, indicating that the momentum space is curved in such a way that its volume remains unchanged regardless how close is the system to the TPT. The wrapping number and the nonlocal topological marker are easily evaluated as 
\begin{eqnarray}
&&{\rm deg}[{\bf n}]=\frac{1}{2}{\rm Sgn}\,M,
\nonumber \\
&&\tilde{F}_{1D}(R)=\frac{1}{2}e^{-|M|R/v}{\rm Sgn}\,M,
\end{eqnarray}
which correctly capture the change of topological invariant at the critical point $M_{c}=0$. The decay length of the nonlocal marker $\xi\equiv v/|M|$ diverges at the critical point, and indicates that the critical exponent defined from\cite{Chen17} $\xi\sim|M|^{-\nu}$ is $\nu=1$. Finally, the geometrical quantities in Eq.~(\ref{Christoffel_Riemann_Ricci}) are not defined in 1D, so we omit the calculation of these quantities.

\subsection{Quantum geometry of 1D lattice Dirac model}

The lattice version of 1D Dirac model is given by Eq.~(\ref{generic_lattice_Dirac}) with $D=1$, i.e.,
\begin{eqnarray}
d_{0}=M+2B-2B\cos ka,\;\;\;d_{1}=A\sin ka.
\end{eqnarray}
which describes many prototype 1D TIs and TSCs. For instance, for the Su-Schrieffer-Heeger model\cite{Su79} of spinless fermions with nearest-neighbor hopping $t$ and alternating hopping $\delta t$ in class BDI, one can identify $M=2\delta t$, $A=t-\delta t$, and $B=(t-\delta t)/2$ by expanding around the gap-closing point $k=\pi$. Kitaev's spinless $p$-wave superconductor\cite{Kitaev01} in class D that contains chemical potential $\mu$, nearest-neighbor hopping $-t$, and nearest-neighbor pairing $\Delta$ near $k=0$ is equivalent to $M=2t-2\mu$, $A=2\Delta$, and $B=-t$. The quantum metric and Jacobian of this model are
\begin{eqnarray}
&&g_{xx}=\frac{N}{8d^{4}}\left[-2AB+A(M+2B)\cos kx\right]^2,
\nonumber \\
&&J_{k}=\frac{1}{d^{2}}\left[-2AB+A(M+2B)\cos kx\right],
\label{1D_lattice_gxx_Jk}
\end{eqnarray}
satisfying Eq.~(\ref{metric_curvature_correspondence}). 



\section{2D topological materials}

Our goal in this section is to study the quantum geometry of generic 2D topological materials in any of the five topologically nontrivial symmetry classes A, AII, C, D, and DIII, described by the Dirac models in Eq.~(\ref{generic_Dirac_models}) with $D=2$. We will discuss the polar and Cartesian coordinate formalisms for 2D systems in the following sections.

\subsection{Quantum geometry of 2D linear Dirac model in polar coordinates}

In this section, we consider 2D linear Dirac models formulated in the polar coordinates $k^{\mu}=(k,\phi)$, where the ${\bf d}$-vector is parametrized by
\begin{eqnarray}
d_{0}=M,\;\;\;d_{1}=vk\cos\phi,\;\;\;d_{2}=vk\sin\phi,
\label{2D_linear_Dirac_model}
\end{eqnarray}
where $v$ is the Fermi momentum. The components of the metric are
\begin{eqnarray}
&&g_{kk}=\frac{NM^{2}v^{2}}{8(M^{2}+v^{2}k^{2})^{2}}=(g^{kk})^{-1},\;\;\;
\nonumber \\
&&g_{\phi\phi}=\frac{Nv^{2}k^{2}}{8(M^{2}+v^{2}k^{2})}=(g^{\phi\phi})^{-1},\;\;\;
\nonumber \\
&&g_{k\phi}=g_{\phi k}=g^{k\phi}=g^{\phi k}=0.
\label{2D_gkk_gtt}
\end{eqnarray}
A problem immediately emerges, namely in the large momentum $k\rightarrow\infty$ away from the HSP $k=0$, the $g_{kk}$ vanishes and hence $g^{kk}$ diverges
\begin{eqnarray}
\lim_{k\rightarrow\infty}g_{kk}=0,\;\;\;\lim_{k\rightarrow\infty}g^{kk}=\infty,
\label{metric_linear_divergence}
\end{eqnarray}
which is simply an artifact of this linear Dirac model that is certainly not true in real materials. Thus one should keep in mind that this linear Dirac model is valid only around $k=0$, which nevertheless is sufficient to describe the quantum geometrical properties near the critical point. For instance, the Jacobian in the polar coordinates is
\begin{eqnarray}
J_{\bf k}=\frac{1}{d^{3}}\varepsilon_{i_{0}i_{1}i_{2}}d_{i_0}\partial_{k}d_{i_1}\partial_{\phi}d_{i_{2}}
=\frac{Mv^{2}k}{\left(M^{2}+v^{2}k^{2}\right)^{3/2}},
\end{eqnarray}
which gives the half-integer wrapping number 
\begin{eqnarray}
{\rm deg}[{\bf n}]=\frac{1}{4\pi}\int_{0}^{\infty} dk\int_{0}^{2\pi}d\phi\,J_{\bf k}=\frac{1}{2}{\rm Sgn}M,
\end{eqnarray}
that correctly captures the change of topological invariant ${\rm deg}[{\bf n}]_{M>0}-{\rm deg}[{\bf n}]_{M<0}=1$ at the critical point $M_{c}=0$ of this continuous model. The Fourier transform of the Jacobian in this linear Dirac model has the following analytical form 
\begin{eqnarray}
&&\tilde{F}_{2D}(R)=\frac{1}{4\pi}\int_{0}^{\infty}dk\int_{0}^{2\pi}d\phi\,J_{\bf k}\,e^{ikR\cos\phi}
\nonumber \\
&&=\frac{1}{2}\int_{0}^{\infty}dk\,J_{0}(kR)\frac{Mv^{2}k}{(M^{2}+v^{2}k^{2})^{3/2}},
\nonumber \\
&&=\frac{1}{2}e^{-R|M|/v}{\rm Sgn}\,M,
\end{eqnarray}
where $J_{0}(kR)$ is the zeroth-order Bessel function of the first kind (see Sec.~6.554 of Ref.~\onlinecite{Gradshteyn14}). This result indicates that the nonlocal topological marker decays exponentially with the correlation length $\xi=v/|M|$ that diverges at the critical point $M_{c}=0$, thereby serving as an appropriate correlation to describe the critical behavior near the TPTs\cite{Chen17,Chen19_universality,Chen19_AMS_review,Molignini23_Chern_marker,Chen23_spin_Chern_marker}.

The determinant of quantum metric in the polar coordinates satisfies the metric-curvature correspondence in Eq.~(\ref{metric_curvature_correspondence}) by
\begin{eqnarray}
|J_{\bf k}|=\left(\frac{8}{N}\right)\sqrt{\det g_{\mu\nu}}=\frac{|M|v^{2}k}{\left(M^{2}+v^{2}k^{2}\right)^{3/2}}.
\end{eqnarray}
Consequently, the volume of the curved momentum space manifold is again independent of the band gap $M$
\begin{eqnarray}
V_{2D}^{\rm lin}=\int_{0}^{\infty} dk\int_{0}^{2\pi}d\phi\sqrt{\det g_{\mu\nu}}=\frac{N\pi}{4},
\end{eqnarray}
which is only determined by the dimension of the Dirac matrices $N$.



\begin{figure*}[ht!]
\begin{center}
\includegraphics[clip=true,width=1.5\columnwidth]{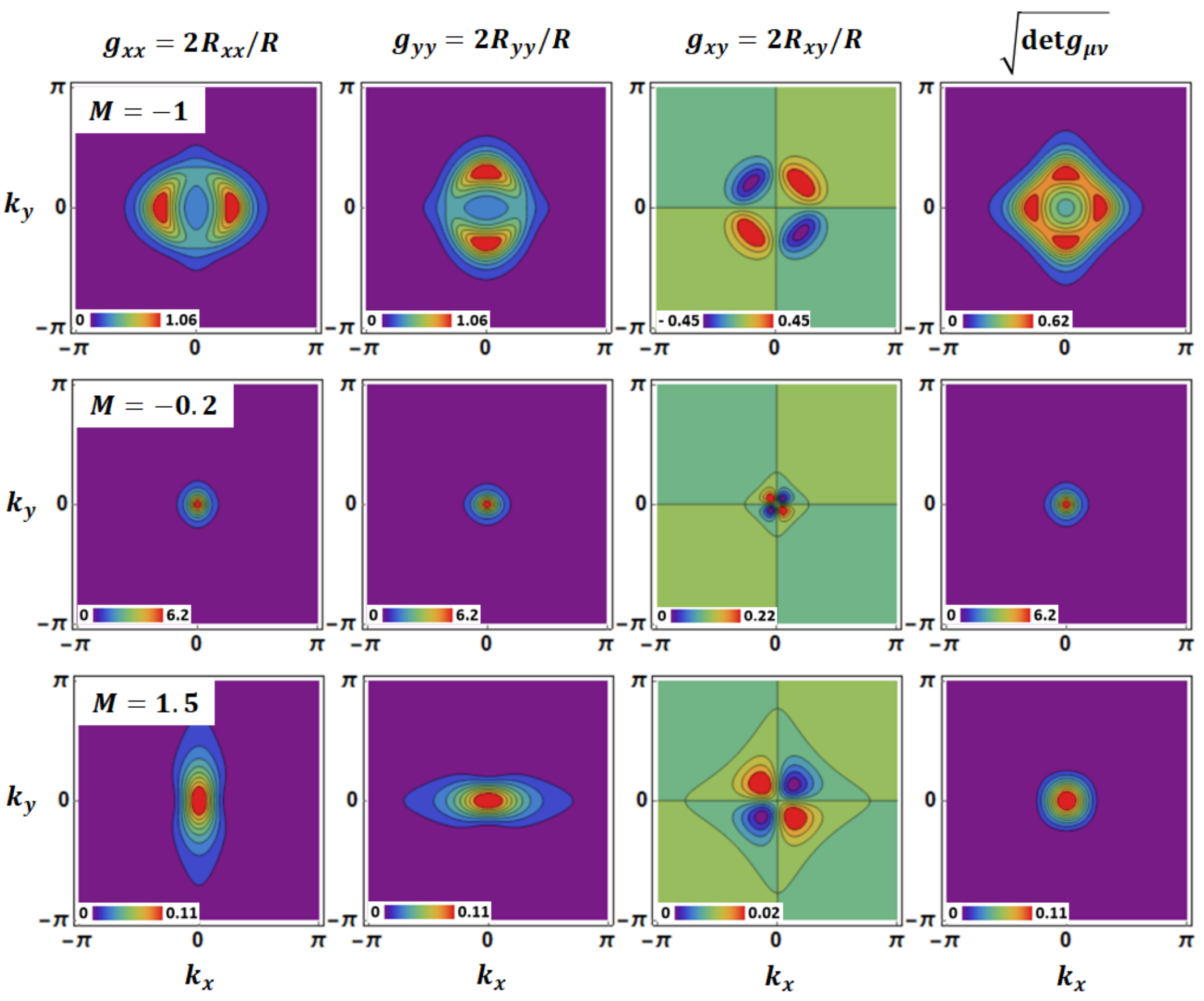}
\caption{Pattern of the quantum metric, or equivalently the rescaled Ricci tensor $g_{\mu\nu}=2R_{\mu\nu}/R$ due to the constant Ricci scalar $R=16/N$, and the volume form $\sqrt{\det g_{\mu\nu}}$ in 2D lattice models of TIs and TSCs plotted as a function of ${\bf k}=(k_{x},k_{y})$. We choose $A=B=1$ in Eq.~(\ref{2D_lattice_Dirac_model}), and the parameters that simulate the topologically nontrivial phase $M=-1$ (top row), approaching the critical point $M=-0.2$ (middle row), and the topologically trivial phase $M=1.5$ (bottom row). } 
\label{fig:2D_gmunu_figure}
\end{center}
\end{figure*}

The derivatives of the metric are
\begin{eqnarray}
&&\partial_{k}g_{kk}=-\frac{NM^{2}v^{4}k}{2(M^{2}+v^{2}k^{2})^{3}},
\nonumber \\
&&\partial_{k}g_{\phi\phi}=\frac{NM^{2}v^{2}k}{4(M^{2}+v^{2}k^{2})^{2}},
\end{eqnarray}
and all other derivatives are zero. Using them to calculate the Christoffel symbol in Eq.~(\ref{Christoffel_Riemann_Ricci}) yields the nonzero components
\begin{eqnarray}
&&\Gamma_{kk}^{k}=-\frac{2v^{2}k}{(M^{2}+v^{2}k^{2})},\;\;\;\Gamma_{\phi\phi}^{k}=-k,
\nonumber \\
&&\Gamma_{k\phi}^{\phi}=\Gamma_{\phi k}^{\phi}=\frac{M^{2}}{k(M^{2}+v^{2}k^{2})},
\end{eqnarray}
from which follows the nonzero components of Riemann tensor and the Ricci tensor
\begin{eqnarray}
&&R^{k}_{\;\phi k\phi}=-R^{k}_{\;\phi\phi k}=\frac{v^{2}k^{2}}{(M^{2}+v^{2}k^{2})},
\nonumber \\
&&R^{\phi}_{\;kk\phi}=-R^{\phi}_{\;k\phi k}=-\frac{M^{2}v^{2}}{(M^{2}+v^{2}k^{2})^{2}},
\nonumber \\
&&R_{\mu\nu}=\frac{8}{N}g_{\mu\nu}.
\end{eqnarray} 
They lead to a Ricci scalar that depends on the dimension $N$ of the Dirac matrices
\begin{eqnarray}
R=\frac{16}{N},
\label{2D_R16N_G0}
\end{eqnarray}
resulting in a vanishing Einstein tensor $G_{\mu\nu}=0$. This   
indicates that the vacuum Einstein equation is satisfied, as it should be for any metric in 2D. Moreover, the constant Ricci scalar implies that the Euler characteristic of this linear Dirac model is always unity
\begin{eqnarray}
\chi=\frac{1}{4\pi}\int_{0}^{\infty} dk\int_{0}^{2\pi}d\phi\sqrt{\det g_{\mu\nu}}\,R=1,
\end{eqnarray}
for both topological phases $M>0$ and $M<0$.

The geodesic equations are explicitly given by
\begin{eqnarray}
&&\frac{d^{2}k}{dt^{2}}-\frac{2v^{2}k}{(M^{2}+v^{2}k^{2})}\left(\frac{dk}{dt}\right)^{2}
-k\left(\frac{d\phi}{dt}\right)^{2}=0,
\nonumber \\
&&\frac{d^{2}\phi}{dt^{2}}+\frac{2M^{2}}{k(M^{2}+v^{2}k^{2})}\left(\frac{dk}{dt}\right)\left(\frac{d\phi}{dt}\right)=0.
\label{geodesic_eq_2D}
\end{eqnarray}
These geodesic equations can be solved in the same way as solving the Kepler problem\cite{Symon71}. First we write $k=k(\phi(t))$ as a function of $\phi(t)$, such that 
\begin{eqnarray}
\frac{dk}{dt}=\frac{dk}{d\phi}\frac{d\phi}{dt},\;\;\;
\frac{d^{2}k}{dt^{2}}=\frac{d^{2}k}{d\phi^{2}}\left(\frac{d\phi}{dt}\right)^{2}+\frac{dk}{d\phi}\frac{d^{2}\phi}{dt^{2}}.
\end{eqnarray}
Applying them to the geodesic equations in Eq.~(\ref{geodesic_eq_2D}), and putting the second geodesic equation to the first one, we obtain
\begin{eqnarray}
k\frac{d^{2}k}{d\phi^{2}}-2\left(\frac{dk}{d\phi}\right)^{2}-k^{2}=0.
\end{eqnarray}
The inverse of the radius $u=1/k$ satisfies a harmonic oscillator equation, and hence the solution is 
\begin{eqnarray}
\frac{d^{2}u}{d\phi^{2}}+u=0,\;\;\;u=\frac{1}{A}\cos\phi,\;\;\;k=A\sec\phi,
\end{eqnarray} 
Thus we see that the geodesic in the Cartesian coordinates is a straight line, since 
\begin{eqnarray}
k_{x}=k\cos\phi=A,\;\;\;k_{y}=k\sin\phi=A\tan\phi.
\end{eqnarray}
where $A$ is determined by the initial condition of the geodesic. Interestingly, in the topological semimetal limit $M\rightarrow 0$, the second equation in Eq.~(\ref{geodesic_eq_2D}) becomes $d^{2}\phi/dt^{2}=0$ and hence $d\phi/dt=v_{\phi}$ is a constant, meaning that a hypothetical observer would move along the geodesic in such a way what the angular velocity $v_{\phi}$ is constant.

\subsection{Quantum geometry of 2D lattice Dirac model in Cartesian coordinates}

We proceed to examine the lattice model of Dirac Hamiltonian constructed from regularizing the linear model in Eq.~(\ref{2D_linear_Dirac_model}) on a square lattice, which is parametrized by
\begin{eqnarray}
&&d_{0}=M+4B-2B\cos k^{x}-2B\cos k^{y},
\nonumber \\
&&d_{1}=A\sin k^{x},\;\;\;d_{2}=A\sin k^{y}.
\label{2D_lattice_Dirac_model}
\end{eqnarray}
This parametrization describes several prototype 2D TIs such as the Chern insulator in class A and Bernevig-Hughes-Zhang model\cite{Bernevig13,Bernevig06} in class AII, as well as 2D chiral and helical $p$-wave TSCs in class D and DIII, respectively\cite{Schnyder08}. The metric in Cartesian coordinates $k^{\mu}=(k^{x},k^{y})$ can be calculated straightforwardly using Eq.~(\ref{quantum_metric_general_formula}), and the geometrical quantitites in Eq.~(\ref{Christoffel_Riemann_Ricci}) follows directly from the derivatives. Their precise expression of these quantities are rather lengthy, so we omit for simplicity. Nevertheless, the Jacobian $J_{\bf k}$ has a rather compact form
\begin{eqnarray}
J_{\bf k}&=&\frac{A^{2}}{d^{3}}\left[-2B\cos k^{x}-2B\cos k^{y}\right.
\nonumber \\
&&\left.+(4B+M)\cos k^{x}\cos k^{y}\right],
\end{eqnarray}
and hence the analytical expression for the volume form $\sqrt{\det g_{\mu\nu}}$ can be easily found from the metric-curvature correspondence in Eq.~(\ref{metric_curvature_correspondence}).
The momentum profile of these quantities can be calculated numerically in a straightforward manner. Remarkably, we find that Eq.~(\ref{2D_R16N_G0}) remains true in the lattice Dirac model, i.e., the Ricci scalar is a constant $R=16/N$ everywhere in the momentum space despite the complicated momentum-dependence of the ${\bf d}$-vector in Eq.~(\ref{2D_lattice_Dirac_model}), and the vacuum Einstein equation is satisfied without a cosmological constant $\Lambda=0$, as it should be.

The numerical results for the quantum metric, or equivalently the rescaled Ricci tensor $g_{\mu\nu}=2R_{\mu\nu}/R$, and the volume form in momentum space are shown in Fig.~\ref{fig:2D_gmunu_figure} for three parameters that simulate the topologically nontrivial phase, close to the critical point (equivalently the topological semimetal), and the topologically trivial phase. We see that the pattern of the metric is very concentrated near the HSP ${\bf k}=(0,0)$ where the TPT takes place. As the system approaches the critical point $M_{c}=0$, the metric and volume form gradually narrows and diverges owing to the critical behavior of the curvature function $J_{\bf k}$. Interestingly, the lobes of $g_{xx}$ extends along $k_{x}$ direction in the nontrivial phase $M<0$, but rotates to be along the $k_{y}$ directions as the system enters the trivial phase $M>0$, and likewisely for the $g_{yy}$, indicating that the lattice model displays a much more complicated momentum profile of $g_{\mu\nu}$ than that of the continuous model described by Eq.~(\ref{2D_gkk_gtt}).

\begin{figure*}[ht!]
\begin{center}
\includegraphics[clip=true,width=1.9\columnwidth]{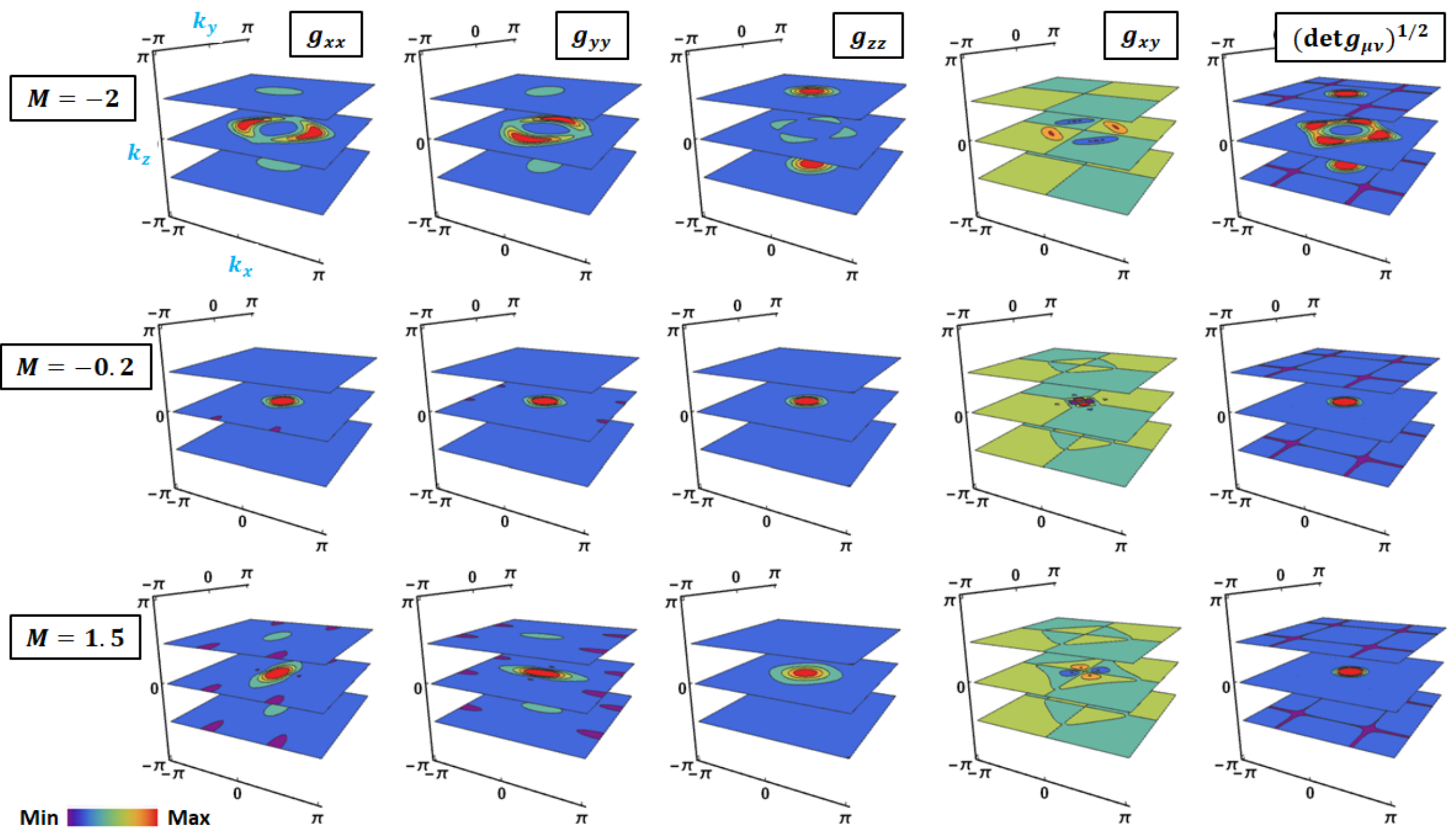}
\caption{Pattern of the quantum metric, equivalently the rescaled Ricci tensor $g_{\mu\nu}=3R_{\mu\nu}/R$ due to the constant Ricci scalar $R=48/N$ and cosmological constant $\Lambda=8/N$, and the volume form $\sqrt{\det g_{\mu\nu}}$ in 3D lattice models of TIs and TSCs plotted as a function of ${\bf k}=(k_{x},k_{y},k_{z})$ (coordinates are labeled at the upper left figure). The parameters in Eq.~(\ref{3D_lattice_Dirac_model}) are $A=B=1$, and we examine topologically nontrivial phase $M=-2$ (top row), approaching the critical point $M=-0.2$ (middle row), and the topologically trivial phase $M=1.5$ (bottom row). In each figure, the red color is the maximum and blue the minimum. } 
\label{fig:3D_gxx_gxy_figure}
\end{center}
\end{figure*}

\section{3D topological materials}

The topologically nontrivial symmetry classes in 3D are classes AII, AIII, CII, DIII, and CI. Once again their quantum metric is all described by Eq.~(\ref{quantum_metric_general_formula}) with a properly chosen $N$, and so are the differential geometrical properties of the 3D momentum space. We will first investigate the linear Dirac model in the spherical coordinates, and then turn to lattice models in Cartesian coordinates.

\subsection{Quantum geometry of 3D linear Dirac model in spherical coordinates}

The 3D linear Dirac model in the spherical coordinates is parametrized by
\begin{eqnarray}
&&d_{0}=M,\;\;\;d_{1}=vk\sin\theta\cos\phi,\;\;\;
\nonumber \\
&&d_{2}=vk\sin\theta\sin\phi,\;\;\;
d_{3}=vk\cos\theta.
\label{3D_linear_Dirac_model}
\end{eqnarray}
The Jacobian in the spherical coordinates is
\begin{eqnarray}
&&J_{\bf k}=\frac{1}{d^{4}}\varepsilon_{i_{0}i_{1}i_{2}i_{3}}d_{i_{0}}
\partial_{k}d_{i_{1}}\partial_{\theta}d_{i_{2}}\partial_{\phi}d_{i_{3}}
\nonumber \\
&&=\frac{Mv^{3}k^{2}\sin\theta}{(M^{2}+v^{2}k^{2})^{2}},
\end{eqnarray}
which gives rise to the topological invariant 
\begin{eqnarray}
{\rm deg}[{\bf n}]=\frac{1}{2\pi^{2}}\int_{0}^{\infty}dk\int_{0}^{\pi}d\theta\int_{0}^{2\pi}d\phi\,J_{\bf k}=\frac{1}{2}{\rm Sgn}M,\;\;\;
\end{eqnarray}
that correctly captures the change of topological order across the transition ${\rm deg}[{\bf n}]_{M>0}-{\rm deg}[{\bf n}]_{M<0}=1$. The Fourier transform of the Jacobian again gives a nonlocal topological marker\cite{Chen23_universal_marker}
\begin{eqnarray}
&&\tilde{F}_{3D}(R)=\frac{1}{2\pi^{2}}\int_{0}^{\infty}dk\int_{0}^{\pi}d\theta\int_{0}^{2\pi}d\phi\,J_{\bf k}\,e^{ikR\cos\theta}
\nonumber \\
&&=\frac{2}{\pi R}\int_{0}^{\infty}dk\,\frac{Mv^{3}k\sin kR}{(M^{2}+v^{2}k^{2})^{2}},
\nonumber \\
&&=\frac{1}{2}e^{-R|M|/v}{\rm Sgn}\,M,
\end{eqnarray}
after a contour integration, indicating that it serves as an appropriate correlation function that decays exponentially with the correlation length $\xi=v/|M|$, thereby characterizes the critical behavior at the TPT $M\rightarrow 0$.




Using Eq.~(\ref{quantum_metric_general_formula}), one obtains the nonzero components of the metric
\begin{eqnarray}
&&g_{kk}=\frac{NM^{2}v^{2}}{8(M^{2}+v^{2}k^{2})^{2}}=(g^{kk})^{-1},\;\;\;
\nonumber \\
&&g_{\theta\theta}=\frac{Nv^{2}k^{2}}{8(M^{2}+v^{2}k^{2})}=(g^{\theta\theta})^{-1},\;\;\;
\nonumber \\
&&g_{\phi\phi}=\frac{Nv^{2}k^{2}\sin^{2}\theta}{8(M^{2}+v^{2}k^{2})}=(g^{\phi\phi})^{-1}.\;\;\;
\end{eqnarray}
Again we see that the vanishing $g_{\mu\nu}$ and diverging $g^{\mu\nu}$ at large $k$ as described by Eq.~(\ref{metric_linear_divergence}), indicating that the linear Dirac model cannot capture the geometrical properties far away from the HSP in real materials. Nevertheless, the model still correctly captures the critical behavior near the TPT. For instance, the metric-curvature correspondence in Eq.~(\ref{metric_curvature_correspondence}) is satisfied by
\begin{eqnarray}
|J_{\bf k}|=\left(\frac{8}{N}\right)^{3/2}\sqrt{\det g_{\mu\nu}}=\frac{|M||\sin\theta|v^{3}k^{2}}{\left(M^{2}+v^{2}k^{2}\right)^{2}}.
\end{eqnarray}
The volume of the curved 3D momentum space manifold is once again independent of the band gap $M$
\begin{eqnarray}
&&V_{3D}^{\rm lin}=\int_{0}^{\infty} dk\int_{0}^{\pi}d\theta\int_{0}^{2\pi}d\phi\sqrt{\det g_{\mu\nu}}
\nonumber \\
&&=\pi^{2}\left(\frac{N}{8}\right)^{3/2}.
\end{eqnarray}
The nonzero Christoffel symbols are 
\begin{eqnarray}
&&\Gamma_{kk}^{k}=-\frac{2v^{2}k}{(M^{2}+v^{2}k^{2})},\;\;\;\Gamma_{\theta\theta}^{k}=-k,\;\;\;
\Gamma_{\phi\phi}^{k}=-k\sin^{2}\theta,\;\;\;
\nonumber \\
&&\Gamma_{k\theta}^{\theta}=\Gamma_{\theta k}^{\theta}=\Gamma_{k\phi}^{\phi}=\Gamma_{\phi k}^{\phi}=\frac{M^{2}}{k(M^{2}+v^{2}k^{2})},
\nonumber \\
&&\Gamma_{\phi\phi}^{\theta}=-\sin\theta\cos\theta,\;\;\;
\Gamma_{\theta\phi}^{\phi}=\Gamma_{\phi\theta}^{\phi}=\cot\theta.
\end{eqnarray}
yielding nonzero components of the Riemann tensor
\begin{eqnarray}
&&R^{k}_{\;\theta k\theta}=R^{\phi}_{\;\theta\phi\theta}=\frac{v^{2}k^{2}}{(M^{2}+v^{2}k^{2})},
\nonumber \\
&&R^{k}_{\;\phi k\phi}=R^{\theta}_{\;\phi \theta\phi}=\frac{v^{2}k^{2}\sin^{2}\theta}{(M^{2}+v^{2}k^{2})},
\nonumber \\
&&R^{\theta}_{\;k\theta k}=R^{\phi}_{\;k\phi k}=\frac{M^{2}v^{2}}{(M^{2}+v^{2}k^{2})^{2}},
\end{eqnarray}
and note that $R^{\rho}_{\;\sigma\mu\nu}=-R^{\rho}_{\;\sigma\nu\mu}$. As a result, the Ricci tensor is proportional to the metric, and the Ricci scalar is a constant that only depending on the dimension $N$ of the Dirac matrices, and hence the Einstein tensor is also proportional to the quantum metric  
\begin{eqnarray}
R_{\mu\nu}=\frac{16}{N}g_{\mu\nu},\;\;\;R=\frac{48}{N},\;\;\;G_{\mu\nu}=-\frac{8}{N}g_{\mu\nu},
\label{3D_Ricci_scalar_Einstein}
\end{eqnarray}
Thus if we write the vacuum Einstein equation as
\begin{eqnarray}
R_{\mu\nu}-\frac{1}{2}Rg_{\mu\nu}+\Lambda g_{\mu\nu}=0,\;\;\;
\Lambda=\frac{8}{N},
\label{3D_cosmological_constant}
\end{eqnarray}
then we see that the cosmological constant takes the value $\Lambda=8/N$ in 3D linear Dirac models.


The geodesic equations in 3D are
\begin{eqnarray}
&&\frac{d^{2}k}{dt^{2}}-\frac{2v^{2}k}{(M^{2}+v^{2}k^{2})}\left(\frac{dk}{dt}\right)^{2}
-k\left(\frac{d\theta}{dt}\right)^{2}
\nonumber \\
&&-k\sin^{2}\theta\left(\frac{d\phi}{dt}\right)^{2}=0,
\nonumber \\
&&\frac{d^{2}\theta}{dt^{2}}+\frac{2M^{2}}{k(M^{2}+v^{2}k^{2})}\left(\frac{dk}{dt}\right)\left(\frac{d\theta}{dt}\right)
\nonumber \\
&&-\sin\theta\cos\theta\left(\frac{d\phi}{dt}\right)^{2}=0,
\nonumber \\
&&\frac{d^{2}\phi}{dt^{2}}+\frac{2M^{2}}{k(M^{2}+v^{2}k^{2})}\left(\frac{dk}{dt}\right)\left(\frac{d\phi}{dt}\right)
\nonumber \\
&&+2\cot\theta\left(\frac{d\theta}{dt}\right)\left(\frac{d\phi}{dt}\right)=0.
\label{geodesic_eq_3D}
\end{eqnarray}
There is no explicit $\phi$ dependence in all the equations, so we may consider the solution $\phi=0$ and $d\phi/dt=0$, meaning that the geodesic lies on the $xz$-plane in the Cartesian coordinates. The geodesic equations in Eq.~(\ref{geodesic_eq_3D}) then becomes the same as that in 2D in Eq.~(\ref{geodesic_eq_2D}), so we arrive at the same conclusion that the geodesic on the $xz$-plane is a straight line. Since the definition of axes of the coordinates is arbitrary owing to the spherical symmetry, we conclude that any geodesic is a straight line for 3D linear Dirac models.

\subsection{Quantum geometry of 3D lattice Dirac model in Cartesian coordinates}

The 3D lattice Dirac model is obtained by regularizing Eq.~(\ref{3D_linear_Dirac_model}) on a cubic lattice, yielding 
\begin{eqnarray}
&&d_{0}=M+6B-2B\cos k^{x}-2B\cos k^{y}-2B\cos k^{z},
\nonumber \\
&&d_{1}=A\sin k^{x},\;\;\;d_{2}=A\sin k^{y},\;\;\;d_{3}=A\sin k^{z},
\label{3D_lattice_Dirac_model}
\end{eqnarray}
which is appropriate to describe prototype 3D TIs in class AII such as\cite{Zhang09,Liu10} Bi$_{2}$Se$_{3}$ and Bi$_{2}$Te$_{3}$, and the B-phase of superfluid $^{3}$He in class DIII\cite{Balian63,Volovik09}. The metric and geometrical quantities can be calculated in the Cartesian coordinates according to Eqs.~(\ref{quantum_metric_general_formula}) and (\ref{Christoffel_Riemann_Ricci}). Although these quantities have a rather lengthy expression that we omit for simplicity, the Jacobian has a rather simple form
\begin{eqnarray}
&&J_{\bf k}=\frac{A^{3}}{d^{4}}\left[-2B\cos k^{x}\cos k^{y}-2B\cos k^{y}\cos k^{z}\right.
\nonumber \\
&&\left.-2B\cos k^{z}\cos k^{x}+(M+6B)\cos k^{x}\cos k^{y}\cos k^{z}\right],\;\;\;
\end{eqnarray}
and hence the volume form $\sqrt{\det g_{\mu\nu}}$ can be easily obtained from Eq.~(\ref{metric_curvature_correspondence}). Numerically, we find that Eqs.~(\ref{3D_Ricci_scalar_Einstein}) and (\ref{3D_cosmological_constant}) remain true in the lattice Dirac model, i.e., the Ricci scalar is still a constant $R=48/N$ throughout the BZ, and the vacuum Einstein equation is satisfied with a cosmological constant $\Lambda=8/N$ despite the complicated momentum-dependence of the ${\bf d}$-vector in Eq.~(\ref{3D_lattice_Dirac_model}).

Numerical results for the quantum metric, equivalently the rescaled Ricci tensor $g_{\mu\nu}=3R_{\mu\nu}/R$, is shown in Fig.~\ref{fig:3D_gxx_gxy_figure}, for three parameters of the mass term $M$ chosen in the topologically nontrivial phase, close to the critical point, and in the trivial phase. The evolution of the metric is similar to that in 2D models shown in Fig.~\ref{fig:2D_gmunu_figure}, i.e., one sees that in the nontrivial phase $M<0$ the diagonal elements of the metric $g_{\mu\mu}$ have lobes along ${\hat{\boldsymbol\mu}}$ direction. These lobes gradually shrink and diverge at the HSP ${\bf k}=(0,0,0)$ as the system approaches the critical point $M\rightarrow 0$. As the system enters the trivial phase $M>0$, the lobes of $g_{xx}$ turn to the $k_{y}$ direction and that of $g_{yy}$ turn to the $k_{x}$ direction, similar to that in 2D, whereas $g_{zz}$ gradually flattens in the $k_{x}k_{y}$-plane. These features of the metric once again indicate that the lattice model has a more complicated behavior of the metric than the continuous model discussed in the previous section.

\section{Conclusions}

In summary, we investigate the differential geometrical properties of the momentum space manifold of TIs and TSCs equipped with the quantum metric. Within the context of Dirac models, the quantum metric of the filled valence band states or quasihole states has a simple analytical form in any dimension and symmetry class, allowing us to systematically study the resulting Christopher symbol, Riemann tensor, Ricci scalar, geodesics, Einstein equation, and a nonlocal topological marker, which can measured in TIs by pump-probe experiments. For 2D linear Dirac models, we find that the quantum metric yields a constant Ricci scalar determined only by the dimension of the Dirac models. The Euler characteristic is found to be always unity, and the geodesic is always a straight line, and the nonlocal topological marker decays exponentially with a correlation length that is inversely proportional to the band gap. For 2D lattice Dirac models, the constant Ricci scalar, zero cosmological constant, and unity Euler characteristic remain true, but the quantum metric and Ricci tensor display a more complicated momentum profiled depending on the mass term of the Dirac model.

For 3D linear Dirac models, the geodesic is still a straight line, the nonlocal marker still exponentially decays, and the Ricci scalar is also a constant, but the cosmological constant is found to be finite and only determined by the dimension of the Dirac Hamiltonian. The constant Ricci scalar and finite cosmological constant are found to remain true even for 3D lattice Dirac models, yet the pattern of quantum metric and Ricci tensor strongly depends on the mass term. Our results thus reveal a number of universal, parameter independent differential geometrical properties that ubiquitously apply to any topological materials described by Dirac models, which helps to understand the quantum geometry in these materials. Whether these peculiar geometrical properties, such as the constant Ricci scalar and cosmological constant, have nontrivial consequences on certain material properties is an interesting subject that awaits to be further explored.

\bibliography{Literatur}

\begin{thebibliography}{52}%
\makeatletter
\providecommand \@ifxundefined [1]{%
 \@ifx{#1\undefined}
}%
\providecommand \@ifnum [1]{%
 \ifnum #1\expandafter \@firstoftwo
 \else \expandafter \@secondoftwo
 \fi
}%
\providecommand \@ifx [1]{%
 \ifx #1\expandafter \@firstoftwo
 \else \expandafter \@secondoftwo
 \fi
}%
\providecommand \natexlab [1]{#1}%
\providecommand \enquote  [1]{``#1''}%
\providecommand \bibnamefont  [1]{#1}%
\providecommand \bibfnamefont [1]{#1}%
\providecommand \citenamefont [1]{#1}%
\providecommand \href@noop [0]{\@secondoftwo}%
\providecommand \href [0]{\begingroup \@sanitize@url \@href}%
\providecommand \@href[1]{\@@startlink{#1}\@@href}%
\providecommand \@@href[1]{\endgroup#1\@@endlink}%
\providecommand \@sanitize@url [0]{\catcode `\\12\catcode `\$12\catcode
  `\&12\catcode `\#12\catcode `\^12\catcode `\_12\catcode `\%12\relax}%
\providecommand \@@startlink[1]{}%
\providecommand \@@endlink[0]{}%
\providecommand \url  [0]{\begingroup\@sanitize@url \@url }%
\providecommand \@url [1]{\endgroup\@href {#1}{\urlprefix }}%
\providecommand \urlprefix  [0]{URL }%
\providecommand \Eprint [0]{\href }%
\providecommand \doibase [0]{http://dx.doi.org/}%
\providecommand \selectlanguage [0]{\@gobble}%
\providecommand \bibinfo  [0]{\@secondoftwo}%
\providecommand \bibfield  [0]{\@secondoftwo}%
\providecommand \translation [1]{[#1]}%
\providecommand \BibitemOpen [0]{}%
\providecommand \bibitemStop [0]{}%
\providecommand \bibitemNoStop [0]{.\EOS\space}%
\providecommand \EOS [0]{\spacefactor3000\relax}%
\providecommand \BibitemShut  [1]{\csname bibitem#1\endcsname}%
\let\auto@bib@innerbib\@empty
\bibitem [{\citenamefont {Hasan}\ and\ \citenamefont {Kane}(2010)}]{Hasan10}%
  \BibitemOpen
  \bibfield  {author} {\bibinfo {author} {\bibfnamefont {M.~Z.}\ \bibnamefont
  {Hasan}}\ and\ \bibinfo {author} {\bibfnamefont {C.~L.}\ \bibnamefont
  {Kane}},\ }\href {\doibase 10.1103/RevModPhys.82.3045} {\bibfield  {journal}
  {\bibinfo  {journal} {Rev. Mod. Phys.}\ }\textbf {\bibinfo {volume} {82}},\
  \bibinfo {pages} {3045} (\bibinfo {year} {2010})}\BibitemShut {NoStop}%
\bibitem [{\citenamefont {Qi}\ and\ \citenamefont {Zhang}(2011)}]{Qi11}%
  \BibitemOpen
  \bibfield  {author} {\bibinfo {author} {\bibfnamefont {X.-L.}\ \bibnamefont
  {Qi}}\ and\ \bibinfo {author} {\bibfnamefont {S.-C.}\ \bibnamefont {Zhang}},\
  }\href {\doibase 10.1103/RevModPhys.83.1057} {\bibfield  {journal} {\bibinfo
  {journal} {Rev. Mod. Phys.}\ }\textbf {\bibinfo {volume} {83}},\ \bibinfo
  {pages} {1057} (\bibinfo {year} {2011})}\BibitemShut {NoStop}%
\bibitem [{\citenamefont {Zak}(1989)}]{Zak89}%
  \BibitemOpen
  \bibfield  {author} {\bibinfo {author} {\bibfnamefont {J.}~\bibnamefont
  {Zak}},\ }\href {\doibase 10.1103/PhysRevLett.62.2747} {\bibfield  {journal}
  {\bibinfo  {journal} {Phys. Rev. Lett.}\ }\textbf {\bibinfo {volume} {62}},\
  \bibinfo {pages} {2747} (\bibinfo {year} {1989})}\BibitemShut {NoStop}%
\bibitem [{\citenamefont {Thouless}\ \emph {et~al.}(1982)\citenamefont
  {Thouless}, \citenamefont {Kohmoto}, \citenamefont {Nightingale},\ and\
  \citenamefont {den Nijs}}]{Thouless82}%
  \BibitemOpen
  \bibfield  {author} {\bibinfo {author} {\bibfnamefont {D.~J.}\ \bibnamefont
  {Thouless}}, \bibinfo {author} {\bibfnamefont {M.}~\bibnamefont {Kohmoto}},
  \bibinfo {author} {\bibfnamefont {M.~P.}\ \bibnamefont {Nightingale}}, \ and\
  \bibinfo {author} {\bibfnamefont {M.}~\bibnamefont {den Nijs}},\ }\href
  {\doibase 10.1103/PhysRevLett.49.405} {\bibfield  {journal} {\bibinfo
  {journal} {Phys. Rev. Lett.}\ }\textbf {\bibinfo {volume} {49}},\ \bibinfo
  {pages} {405} (\bibinfo {year} {1982})}\BibitemShut {NoStop}%
\bibitem [{\citenamefont {Berry}(1984)}]{Berry84}%
  \BibitemOpen
  \bibfield  {author} {\bibinfo {author} {\bibfnamefont {M.~V.}\ \bibnamefont
  {Berry}},\ }\href {\doibase 10.1098/rspa.1984.0023} {\bibfield  {journal}
  {\bibinfo  {journal} {Proc. R. Soc. A}\ }\textbf {\bibinfo {volume} {392}},\
  \bibinfo {pages} {45} (\bibinfo {year} {1984})}\BibitemShut {NoStop}%
\bibitem [{\citenamefont {Provost}\ and\ \citenamefont
  {Vallee}(1980)}]{Provost80}%
  \BibitemOpen
  \bibfield  {author} {\bibinfo {author} {\bibfnamefont {J.~P.}\ \bibnamefont
  {Provost}}\ and\ \bibinfo {author} {\bibfnamefont {G.}~\bibnamefont
  {Vallee}},\ }\href {https://projecteuclid.org:443/euclid.cmp/1103908308}
  {\bibfield  {journal} {\bibinfo  {journal} {Comm. Math. Phys.}\ }\textbf
  {\bibinfo {volume} {76}},\ \bibinfo {pages} {289} (\bibinfo {year}
  {1980})}\BibitemShut {NoStop}%
\bibitem [{\citenamefont {You}\ \emph {et~al.}(2007)\citenamefont {You},
  \citenamefont {Li},\ and\ \citenamefont {Gu}}]{You07}%
  \BibitemOpen
  \bibfield  {author} {\bibinfo {author} {\bibfnamefont {W.-L.}\ \bibnamefont
  {You}}, \bibinfo {author} {\bibfnamefont {Y.-W.}\ \bibnamefont {Li}}, \ and\
  \bibinfo {author} {\bibfnamefont {S.-J.}\ \bibnamefont {Gu}},\ }\href
  {\doibase 10.1103/PhysRevE.76.022101} {\bibfield  {journal} {\bibinfo
  {journal} {Phys. Rev. E}\ }\textbf {\bibinfo {volume} {76}},\ \bibinfo
  {pages} {022101} (\bibinfo {year} {2007})}\BibitemShut {NoStop}%
\bibitem [{\citenamefont {Zanardi}\ \emph {et~al.}(2007)\citenamefont
  {Zanardi}, \citenamefont {Giorda},\ and\ \citenamefont
  {Cozzini}}]{Zanardi07}%
  \BibitemOpen
  \bibfield  {author} {\bibinfo {author} {\bibfnamefont {P.}~\bibnamefont
  {Zanardi}}, \bibinfo {author} {\bibfnamefont {P.}~\bibnamefont {Giorda}}, \
  and\ \bibinfo {author} {\bibfnamefont {M.}~\bibnamefont {Cozzini}},\ }\href
  {\doibase 10.1103/PhysRevLett.99.100603} {\bibfield  {journal} {\bibinfo
  {journal} {Phys. Rev. Lett.}\ }\textbf {\bibinfo {volume} {99}},\ \bibinfo
  {pages} {100603} (\bibinfo {year} {2007})}\BibitemShut {NoStop}%
\bibitem [{\citenamefont {Gu}\ \emph {et~al.}(2008)\citenamefont {Gu},
  \citenamefont {Kwok}, \citenamefont {Ning},\ and\ \citenamefont
  {Lin}}]{Gu08}%
  \BibitemOpen
  \bibfield  {author} {\bibinfo {author} {\bibfnamefont {S.-J.}\ \bibnamefont
  {Gu}}, \bibinfo {author} {\bibfnamefont {H.-M.}\ \bibnamefont {Kwok}},
  \bibinfo {author} {\bibfnamefont {W.-Q.}\ \bibnamefont {Ning}}, \ and\
  \bibinfo {author} {\bibfnamefont {H.-Q.}\ \bibnamefont {Lin}},\ }\href
  {\doibase 10.1103/PhysRevB.77.245109} {\bibfield  {journal} {\bibinfo
  {journal} {Phys. Rev. B}\ }\textbf {\bibinfo {volume} {77}},\ \bibinfo
  {pages} {245109} (\bibinfo {year} {2008})}\BibitemShut {NoStop}%
\bibitem [{\citenamefont {Yang}\ \emph {et~al.}(2008)\citenamefont {Yang},
  \citenamefont {Gu}, \citenamefont {Sun},\ and\ \citenamefont {Lin}}]{Yang08}%
  \BibitemOpen
  \bibfield  {author} {\bibinfo {author} {\bibfnamefont {S.}~\bibnamefont
  {Yang}}, \bibinfo {author} {\bibfnamefont {S.-J.}\ \bibnamefont {Gu}},
  \bibinfo {author} {\bibfnamefont {C.-P.}\ \bibnamefont {Sun}}, \ and\
  \bibinfo {author} {\bibfnamefont {H.-Q.}\ \bibnamefont {Lin}},\ }\href
  {\doibase 10.1103/PhysRevA.78.012304} {\bibfield  {journal} {\bibinfo
  {journal} {Phys. Rev. A}\ }\textbf {\bibinfo {volume} {78}},\ \bibinfo
  {pages} {012304} (\bibinfo {year} {2008})}\BibitemShut {NoStop}%
\bibitem [{\citenamefont {Albuquerque}\ \emph {et~al.}(2010)\citenamefont
  {Albuquerque}, \citenamefont {Alet}, \citenamefont {Sire},\ and\
  \citenamefont {Capponi}}]{Albuquerque10}%
  \BibitemOpen
  \bibfield  {author} {\bibinfo {author} {\bibfnamefont {A.~F.}\ \bibnamefont
  {Albuquerque}}, \bibinfo {author} {\bibfnamefont {F.}~\bibnamefont {Alet}},
  \bibinfo {author} {\bibfnamefont {C.}~\bibnamefont {Sire}}, \ and\ \bibinfo
  {author} {\bibfnamefont {S.}~\bibnamefont {Capponi}},\ }\href {\doibase
  10.1103/PhysRevB.81.064418} {\bibfield  {journal} {\bibinfo  {journal} {Phys.
  Rev. B}\ }\textbf {\bibinfo {volume} {81}},\ \bibinfo {pages} {064418}
  (\bibinfo {year} {2010})}\BibitemShut {NoStop}%
\bibitem [{\citenamefont {Gu}(2010)}]{Gu10}%
  \BibitemOpen
  \bibfield  {author} {\bibinfo {author} {\bibfnamefont {S.-J.}\ \bibnamefont
  {Gu}},\ }\href {\doibase 10.1142/S0217979210056335} {\bibfield  {journal}
  {\bibinfo  {journal} {Int. J. Mod. Phys. B}\ }\textbf {\bibinfo {volume}
  {24}},\ \bibinfo {pages} {4371} (\bibinfo {year} {2010})}\BibitemShut
  {NoStop}%
\bibitem [{\citenamefont {Carollo}\ \emph {et~al.}(2020)\citenamefont
  {Carollo}, \citenamefont {Valenti},\ and\ \citenamefont
  {Spagnolo}}]{Carollo20}%
  \BibitemOpen
  \bibfield  {author} {\bibinfo {author} {\bibfnamefont {A.}~\bibnamefont
  {Carollo}}, \bibinfo {author} {\bibfnamefont {D.}~\bibnamefont {Valenti}}, \
  and\ \bibinfo {author} {\bibfnamefont {B.}~\bibnamefont {Spagnolo}},\ }\href
  {\doibase https://doi.org/10.1016/j.physrep.2019.11.002} {\bibfield
  {journal} {\bibinfo  {journal} {Phys. Rep.}\ }\textbf {\bibinfo {volume}
  {838}},\ \bibinfo {pages} {1 } (\bibinfo {year} {2020})},\ \bibinfo {note}
  {geometry of quantum phase transitions}\BibitemShut {NoStop}%
\bibitem [{\citenamefont {{Ma, Yu-Quan}}\ \emph {et~al.}(2013)\citenamefont
  {{Ma, Yu-Quan}}, \citenamefont {{Gu, Shi-Jian}}, \citenamefont {{Chen, Shu}},
  \citenamefont {{Fan, Heng}},\ and\ \citenamefont {{Liu, Wu-Ming}}}]{Ma13}%
  \BibitemOpen
  \bibfield  {author} {\bibinfo {author} {\bibnamefont {{Ma, Yu-Quan}}},
  \bibinfo {author} {\bibnamefont {{Gu, Shi-Jian}}}, \bibinfo {author}
  {\bibnamefont {{Chen, Shu}}}, \bibinfo {author} {\bibnamefont {{Fan, Heng}}},
  \ and\ \bibinfo {author} {\bibnamefont {{Liu, Wu-Ming}}},\ }\href {\doibase
  10.1209/0295-5075/103/10008} {\bibfield  {journal} {\bibinfo  {journal}
  {EPL}\ }\textbf {\bibinfo {volume} {103}},\ \bibinfo {pages} {10008}
  (\bibinfo {year} {2013})}\BibitemShut {NoStop}%
\bibitem [{\citenamefont {Ma}(2014)}]{Ma14}%
  \BibitemOpen
  \bibfield  {author} {\bibinfo {author} {\bibfnamefont {Y.-Q.}\ \bibnamefont
  {Ma}},\ }\href {\doibase 10.1103/PhysRevE.90.042133} {\bibfield  {journal}
  {\bibinfo  {journal} {Phys. Rev. E}\ }\textbf {\bibinfo {volume} {90}},\
  \bibinfo {pages} {042133} (\bibinfo {year} {2014})}\BibitemShut {NoStop}%
\bibitem [{\citenamefont {Yang}\ \emph {et~al.}(2015)\citenamefont {Yang},
  \citenamefont {Ma},\ and\ \citenamefont {Li}}]{Yang15}%
  \BibitemOpen
  \bibfield  {author} {\bibinfo {author} {\bibfnamefont {L.}~\bibnamefont
  {Yang}}, \bibinfo {author} {\bibfnamefont {Y.-Q.}\ \bibnamefont {Ma}}, \ and\
  \bibinfo {author} {\bibfnamefont {X.-G.}\ \bibnamefont {Li}},\ }\href
  {\doibase https://doi.org/10.1016/j.physb.2014.09.022} {\bibfield  {journal}
  {\bibinfo  {journal} {Physica B Condens. Matter}\ }\textbf {\bibinfo {volume}
  {456}},\ \bibinfo {pages} {359} (\bibinfo {year} {2015})}\BibitemShut
  {NoStop}%
\bibitem [{\citenamefont {Pi\'echon}\ \emph {et~al.}(2016)\citenamefont
  {Pi\'echon}, \citenamefont {Raoux}, \citenamefont {Fuchs},\ and\
  \citenamefont {Montambaux}}]{Piechon16}%
  \BibitemOpen
  \bibfield  {author} {\bibinfo {author} {\bibfnamefont {F.}~\bibnamefont
  {Pi\'echon}}, \bibinfo {author} {\bibfnamefont {A.}~\bibnamefont {Raoux}},
  \bibinfo {author} {\bibfnamefont {J.-N.}\ \bibnamefont {Fuchs}}, \ and\
  \bibinfo {author} {\bibfnamefont {G.}~\bibnamefont {Montambaux}},\ }\href
  {\doibase 10.1103/PhysRevB.94.134423} {\bibfield  {journal} {\bibinfo
  {journal} {Phys. Rev. B}\ }\textbf {\bibinfo {volume} {94}},\ \bibinfo
  {pages} {134423} (\bibinfo {year} {2016})}\BibitemShut {NoStop}%
\bibitem [{\citenamefont {Panahiyan}\ \emph {et~al.}(2020)\citenamefont
  {Panahiyan}, \citenamefont {Chen},\ and\ \citenamefont
  {Fritzsche}}]{Panahiyan20_fidelity}%
  \BibitemOpen
  \bibfield  {author} {\bibinfo {author} {\bibfnamefont {S.}~\bibnamefont
  {Panahiyan}}, \bibinfo {author} {\bibfnamefont {W.}~\bibnamefont {Chen}}, \
  and\ \bibinfo {author} {\bibfnamefont {S.}~\bibnamefont {Fritzsche}},\ }\href
  {\doibase 10.1103/PhysRevB.102.134111} {\bibfield  {journal} {\bibinfo
  {journal} {Phys. Rev. B}\ }\textbf {\bibinfo {volume} {102}},\ \bibinfo
  {pages} {134111} (\bibinfo {year} {2020})}\BibitemShut {NoStop}%
\bibitem [{\citenamefont {Schnyder}\ \emph {et~al.}(2008)\citenamefont
  {Schnyder}, \citenamefont {Ryu}, \citenamefont {Furusaki},\ and\
  \citenamefont {Ludwig}}]{Schnyder08}%
  \BibitemOpen
  \bibfield  {author} {\bibinfo {author} {\bibfnamefont {A.~P.}\ \bibnamefont
  {Schnyder}}, \bibinfo {author} {\bibfnamefont {S.}~\bibnamefont {Ryu}},
  \bibinfo {author} {\bibfnamefont {A.}~\bibnamefont {Furusaki}}, \ and\
  \bibinfo {author} {\bibfnamefont {A.~W.~W.}\ \bibnamefont {Ludwig}},\ }\href
  {\doibase 10.1103/PhysRevB.78.195125} {\bibfield  {journal} {\bibinfo
  {journal} {Phys. Rev. B}\ }\textbf {\bibinfo {volume} {78}},\ \bibinfo
  {pages} {195125} (\bibinfo {year} {2008})}\BibitemShut {NoStop}%
\bibitem [{\citenamefont {Ryu}\ \emph {et~al.}(2010)\citenamefont {Ryu},
  \citenamefont {Schnyder}, \citenamefont {Furusaki},\ and\ \citenamefont
  {Ludwig}}]{Ryu10}%
  \BibitemOpen
  \bibfield  {author} {\bibinfo {author} {\bibfnamefont {S.}~\bibnamefont
  {Ryu}}, \bibinfo {author} {\bibfnamefont {A.~P.}\ \bibnamefont {Schnyder}},
  \bibinfo {author} {\bibfnamefont {A.}~\bibnamefont {Furusaki}}, \ and\
  \bibinfo {author} {\bibfnamefont {A.~W.~W.}\ \bibnamefont {Ludwig}},\ }\href
  {http://stacks.iop.org/1367-2630/12/i=6/a=065010} {\bibfield  {journal}
  {\bibinfo  {journal} {New J. Phys.}\ }\textbf {\bibinfo {volume} {12}},\
  \bibinfo {pages} {065010} (\bibinfo {year} {2010})}\BibitemShut {NoStop}%
\bibitem [{\citenamefont {Kitaev}(2009)}]{Kitaev09}%
  \BibitemOpen
  \bibfield  {author} {\bibinfo {author} {\bibfnamefont {A.}~\bibnamefont
  {Kitaev}},\ }\href {\doibase 10.1063/1.3149495} {\bibfield  {journal}
  {\bibinfo  {journal} {AIP Conf. Proc.}\ }\textbf {\bibinfo {volume} {1134}},\
  \bibinfo {pages} {22} (\bibinfo {year} {2009})}\BibitemShut {NoStop}%
\bibitem [{\citenamefont {Chiu}\ \emph {et~al.}(2016)\citenamefont {Chiu},
  \citenamefont {Teo}, \citenamefont {Schnyder},\ and\ \citenamefont
  {Ryu}}]{Chiu16}%
  \BibitemOpen
  \bibfield  {author} {\bibinfo {author} {\bibfnamefont {C.-K.}\ \bibnamefont
  {Chiu}}, \bibinfo {author} {\bibfnamefont {J.~C.~Y.}\ \bibnamefont {Teo}},
  \bibinfo {author} {\bibfnamefont {A.~P.}\ \bibnamefont {Schnyder}}, \ and\
  \bibinfo {author} {\bibfnamefont {S.}~\bibnamefont {Ryu}},\ }\href {\doibase
  10.1103/RevModPhys.88.035005} {\bibfield  {journal} {\bibinfo  {journal}
  {Rev. Mod. Phys.}\ }\textbf {\bibinfo {volume} {88}},\ \bibinfo {pages}
  {035005} (\bibinfo {year} {2016})}\BibitemShut {NoStop}%
\bibitem [{\citenamefont {von Gersdorff}\ \emph {et~al.}(2021)\citenamefont
  {von Gersdorff}, \citenamefont {Panahiyan},\ and\ \citenamefont
  {Chen}}]{vonGersdorff21_unification}%
  \BibitemOpen
  \bibfield  {author} {\bibinfo {author} {\bibfnamefont {G.}~\bibnamefont {von
  Gersdorff}}, \bibinfo {author} {\bibfnamefont {S.}~\bibnamefont {Panahiyan}},
  \ and\ \bibinfo {author} {\bibfnamefont {W.}~\bibnamefont {Chen}},\ }\href
  {\doibase 10.1103/PhysRevB.103.245146} {\bibfield  {journal} {\bibinfo
  {journal} {Phys. Rev. B}\ }\textbf {\bibinfo {volume} {103}},\ \bibinfo
  {pages} {245146} (\bibinfo {year} {2021})}\BibitemShut {NoStop}%
\bibitem [{\citenamefont {von Gersdorff}\ and\ \citenamefont
  {Chen}(2021)}]{vonGersdorff21_metric_curvature}%
  \BibitemOpen
  \bibfield  {author} {\bibinfo {author} {\bibfnamefont {G.}~\bibnamefont {von
  Gersdorff}}\ and\ \bibinfo {author} {\bibfnamefont {W.}~\bibnamefont
  {Chen}},\ }\href {\doibase 10.1103/PhysRevB.104.195133} {\bibfield  {journal}
  {\bibinfo  {journal} {Phys. Rev. B}\ }\textbf {\bibinfo {volume} {104}},\
  \bibinfo {pages} {195133} (\bibinfo {year} {2021})}\BibitemShut {NoStop}%
\bibitem [{\citenamefont {Mera}\ \emph {et~al.}(2022)\citenamefont {Mera},
  \citenamefont {Zhang},\ and\ \citenamefont {Goldman}}]{Mera22}%
  \BibitemOpen
  \bibfield  {author} {\bibinfo {author} {\bibfnamefont {B.}~\bibnamefont
  {Mera}}, \bibinfo {author} {\bibfnamefont {A.}~\bibnamefont {Zhang}}, \ and\
  \bibinfo {author} {\bibfnamefont {N.}~\bibnamefont {Goldman}},\ }\href
  {\doibase 10.21468/SciPostPhys.12.1.018} {\bibfield  {journal} {\bibinfo
  {journal} {SciPost Phys.}\ }\textbf {\bibinfo {volume} {12}},\ \bibinfo
  {pages} {018} (\bibinfo {year} {2022})}\BibitemShut {NoStop}%
\bibitem [{\citenamefont {Kolodrubetz}\ \emph {et~al.}(2013)\citenamefont
  {Kolodrubetz}, \citenamefont {Gritsev},\ and\ \citenamefont
  {Polkovnikov}}]{Kolodrubetz13}%
  \BibitemOpen
  \bibfield  {author} {\bibinfo {author} {\bibfnamefont {M.}~\bibnamefont
  {Kolodrubetz}}, \bibinfo {author} {\bibfnamefont {V.}~\bibnamefont
  {Gritsev}}, \ and\ \bibinfo {author} {\bibfnamefont {A.}~\bibnamefont
  {Polkovnikov}},\ }\href {\doibase 10.1103/PhysRevB.88.064304} {\bibfield
  {journal} {\bibinfo  {journal} {Phys. Rev. B}\ }\textbf {\bibinfo {volume}
  {88}},\ \bibinfo {pages} {064304} (\bibinfo {year} {2013})}\BibitemShut
  {NoStop}%
\bibitem [{\citenamefont {Kolodrubetz}\ \emph {et~al.}(2017)\citenamefont
  {Kolodrubetz}, \citenamefont {Sels}, \citenamefont {Mehta},\ and\
  \citenamefont {Polkovnikov}}]{Kolodrubetz17}%
  \BibitemOpen
  \bibfield  {author} {\bibinfo {author} {\bibfnamefont {M.}~\bibnamefont
  {Kolodrubetz}}, \bibinfo {author} {\bibfnamefont {D.}~\bibnamefont {Sels}},
  \bibinfo {author} {\bibfnamefont {P.}~\bibnamefont {Mehta}}, \ and\ \bibinfo
  {author} {\bibfnamefont {A.}~\bibnamefont {Polkovnikov}},\ }\href {\doibase
  https://doi.org/10.1016/j.physrep.2017.07.001} {\bibfield  {journal}
  {\bibinfo  {journal} {Phys. Rep.}\ }\textbf {\bibinfo {volume} {697}},\
  \bibinfo {pages} {1 } (\bibinfo {year} {2017})}\BibitemShut {NoStop}%
\bibitem [{\citenamefont {Smith}\ \emph {et~al.}(2022)\citenamefont {Smith},
  \citenamefont {Pullasseri},\ and\ \citenamefont {Srivastava}}]{Smith22}%
  \BibitemOpen
  \bibfield  {author} {\bibinfo {author} {\bibfnamefont {T.~B.}\ \bibnamefont
  {Smith}}, \bibinfo {author} {\bibfnamefont {L.}~\bibnamefont {Pullasseri}}, \
  and\ \bibinfo {author} {\bibfnamefont {A.}~\bibnamefont {Srivastava}},\
  }\href {\doibase 10.1103/PhysRevResearch.4.013217} {\bibfield  {journal}
  {\bibinfo  {journal} {Phys. Rev. Res.}\ }\textbf {\bibinfo {volume} {4}},\
  \bibinfo {pages} {013217} (\bibinfo {year} {2022})}\BibitemShut {NoStop}%
\bibitem [{\citenamefont {Chen}(2023{\natexlab{a}})}]{Chen23_universal_marker}%
  \BibitemOpen
  \bibfield  {author} {\bibinfo {author} {\bibfnamefont {W.}~\bibnamefont
  {Chen}},\ }\href {\doibase 10.1103/PhysRevB.107.045111} {\bibfield  {journal}
  {\bibinfo  {journal} {Phys. Rev. B}\ }\textbf {\bibinfo {volume} {107}},\
  \bibinfo {pages} {045111} (\bibinfo {year} {2023}{\natexlab{a}})}\BibitemShut
  {NoStop}%
\bibitem [{\citenamefont {Chen}\ \emph {et~al.}(2017)\citenamefont {Chen},
  \citenamefont {Legner}, \citenamefont {R\"uegg},\ and\ \citenamefont
  {Sigrist}}]{Chen17}%
  \BibitemOpen
  \bibfield  {author} {\bibinfo {author} {\bibfnamefont {W.}~\bibnamefont
  {Chen}}, \bibinfo {author} {\bibfnamefont {M.}~\bibnamefont {Legner}},
  \bibinfo {author} {\bibfnamefont {A.}~\bibnamefont {R\"uegg}}, \ and\
  \bibinfo {author} {\bibfnamefont {M.}~\bibnamefont {Sigrist}},\ }\href
  {\doibase 10.1103/PhysRevB.95.075116} {\bibfield  {journal} {\bibinfo
  {journal} {Phys. Rev. B}\ }\textbf {\bibinfo {volume} {95}},\ \bibinfo
  {pages} {075116} (\bibinfo {year} {2017})}\BibitemShut {NoStop}%
\bibitem [{\citenamefont {Chen}\ and\ \citenamefont
  {Schnyder}(2019)}]{Chen19_universality}%
  \BibitemOpen
  \bibfield  {author} {\bibinfo {author} {\bibfnamefont {W.}~\bibnamefont
  {Chen}}\ and\ \bibinfo {author} {\bibfnamefont {A.~P.}\ \bibnamefont
  {Schnyder}},\ }\href {\doibase 10.1088/1367-2630/ab2a2d} {\bibfield
  {journal} {\bibinfo  {journal} {New J. Phys.}\ }\textbf {\bibinfo {volume}
  {21}},\ \bibinfo {pages} {073003} (\bibinfo {year} {2019})}\BibitemShut
  {NoStop}%
\bibitem [{\citenamefont {Chen}\ and\ \citenamefont
  {Sigrist}(2019)}]{Chen19_AMS_review}%
  \BibitemOpen
  \bibfield  {author} {\bibinfo {author} {\bibfnamefont {W.}~\bibnamefont
  {Chen}}\ and\ \bibinfo {author} {\bibfnamefont {M.}~\bibnamefont {Sigrist}},\
  }\href@noop {} {\emph {\bibinfo {title} {Advanced Topological Insulators, Ch.
  7}}}\ (\bibinfo  {publisher} {Wiley-Scrivener},\ \bibinfo {year}
  {2019})\BibitemShut {NoStop}%
\bibitem [{\citenamefont {Molignini}\ \emph {et~al.}(2023)\citenamefont
  {Molignini}, \citenamefont {Lapierre}, \citenamefont {Chitra},\ and\
  \citenamefont {Chen}}]{Molignini23_Chern_marker}%
  \BibitemOpen
  \bibfield  {author} {\bibinfo {author} {\bibfnamefont {P.}~\bibnamefont
  {Molignini}}, \bibinfo {author} {\bibfnamefont {B.}~\bibnamefont {Lapierre}},
  \bibinfo {author} {\bibfnamefont {R.}~\bibnamefont {Chitra}}, \ and\ \bibinfo
  {author} {\bibfnamefont {W.}~\bibnamefont {Chen}},\ }\href {\doibase
  10.21468/SciPostPhysCore.6.3.059} {\bibfield  {journal} {\bibinfo  {journal}
  {SciPost Phys. Core}\ }\textbf {\bibinfo {volume} {6}},\ \bibinfo {pages}
  {059} (\bibinfo {year} {2023})}\BibitemShut {NoStop}%
\bibitem [{\citenamefont
  {Chen}(2023{\natexlab{b}})}]{Chen23_spin_Chern_marker}%
  \BibitemOpen
  \bibfield  {author} {\bibinfo {author} {\bibfnamefont {W.}~\bibnamefont
  {Chen}},\ }\href {\doibase 10.1088/1361-648X/acba72} {\bibfield  {journal}
  {\bibinfo  {journal} {J. Phys.: Condens. Matter}\ }\textbf {\bibinfo {volume}
  {35}},\ \bibinfo {pages} {155601} (\bibinfo {year}
  {2023}{\natexlab{b}})}\BibitemShut {NoStop}%
\bibitem [{\citenamefont {Carroll}(2003)}]{Carroll03}%
  \BibitemOpen
  \bibfield  {author} {\bibinfo {author} {\bibfnamefont {S.~M.}\ \bibnamefont
  {Carroll}},\ }\href@noop {} {\emph {\bibinfo {title} {Spacetime and Geometry:
  An Introduction to General Relativity}}}\ (\bibinfo  {publisher}
  {Addison-Wesley},\ \bibinfo {year} {2003})\BibitemShut {NoStop}%
\bibitem [{\citenamefont {de~Sousa}\ \emph {et~al.}(2023)\citenamefont
  {de~Sousa}, \citenamefont {Cruz},\ and\ \citenamefont
  {Chen}}]{deSousa23_fidelity_marker}%
  \BibitemOpen
  \bibfield  {author} {\bibinfo {author} {\bibfnamefont {M.~S.~M.}\
  \bibnamefont {de~Sousa}}, \bibinfo {author} {\bibfnamefont {A.~L.}\
  \bibnamefont {Cruz}}, \ and\ \bibinfo {author} {\bibfnamefont
  {W.}~\bibnamefont {Chen}},\ }\href {\doibase 10.1103/PhysRevB.107.205133}
  {\bibfield  {journal} {\bibinfo  {journal} {Phys. Rev. B}\ }\textbf {\bibinfo
  {volume} {107}},\ \bibinfo {pages} {205133} (\bibinfo {year}
  {2023})}\BibitemShut {NoStop}%
\bibitem [{\citenamefont {Souza}\ \emph {et~al.}(2000)\citenamefont {Souza},
  \citenamefont {Wilkens},\ and\ \citenamefont {Martin}}]{Souza00}%
  \BibitemOpen
  \bibfield  {author} {\bibinfo {author} {\bibfnamefont {I.}~\bibnamefont
  {Souza}}, \bibinfo {author} {\bibfnamefont {T.}~\bibnamefont {Wilkens}}, \
  and\ \bibinfo {author} {\bibfnamefont {R.~M.}\ \bibnamefont {Martin}},\
  }\href {\doibase 10.1103/PhysRevB.62.1666} {\bibfield  {journal} {\bibinfo
  {journal} {Phys. Rev. B}\ }\textbf {\bibinfo {volume} {62}},\ \bibinfo
  {pages} {1666} (\bibinfo {year} {2000})}\BibitemShut {NoStop}%
\bibitem [{\citenamefont {Marzari}\ and\ \citenamefont
  {Vanderbilt}(1997)}]{Marzari97}%
  \BibitemOpen
  \bibfield  {author} {\bibinfo {author} {\bibfnamefont {N.}~\bibnamefont
  {Marzari}}\ and\ \bibinfo {author} {\bibfnamefont {D.}~\bibnamefont
  {Vanderbilt}},\ }\href {\doibase 10.1103/PhysRevB.56.12847} {\bibfield
  {journal} {\bibinfo  {journal} {Phys. Rev. B}\ }\textbf {\bibinfo {volume}
  {56}},\ \bibinfo {pages} {12847} (\bibinfo {year} {1997})}\BibitemShut
  {NoStop}%
\bibitem [{\citenamefont {Marzari}\ \emph {et~al.}(2012)\citenamefont
  {Marzari}, \citenamefont {Mostofi}, \citenamefont {Yates}, \citenamefont
  {Souza},\ and\ \citenamefont {Vanderbilt}}]{Marzari12}%
  \BibitemOpen
  \bibfield  {author} {\bibinfo {author} {\bibfnamefont {N.}~\bibnamefont
  {Marzari}}, \bibinfo {author} {\bibfnamefont {A.~A.}\ \bibnamefont
  {Mostofi}}, \bibinfo {author} {\bibfnamefont {J.~R.}\ \bibnamefont {Yates}},
  \bibinfo {author} {\bibfnamefont {I.}~\bibnamefont {Souza}}, \ and\ \bibinfo
  {author} {\bibfnamefont {D.}~\bibnamefont {Vanderbilt}},\ }\href {\doibase
  10.1103/RevModPhys.84.1419} {\bibfield  {journal} {\bibinfo  {journal} {Rev.
  Mod. Phys.}\ }\textbf {\bibinfo {volume} {84}},\ \bibinfo {pages} {1419}
  (\bibinfo {year} {2012})}\BibitemShut {NoStop}%
\bibitem [{\citenamefont {C\'{a}rdenas-Castillo}\ \emph {et~al.}()\citenamefont
  {C\'{a}rdenas-Castillo}, \citenamefont {Zhang}, \citenamefont {Kochan},
  \citenamefont {Freire~Jr.},\ and\ \citenamefont
  {Chen}}]{CardenasCastillo24_spread_Wannier}%
  \BibitemOpen
  \bibfield  {author} {\bibinfo {author} {\bibfnamefont {L.~F.}\ \bibnamefont
  {C\'{a}rdenas-Castillo}}, \bibinfo {author} {\bibfnamefont {S.}~\bibnamefont
  {Zhang}}, \bibinfo {author} {\bibfnamefont {D.}~\bibnamefont {Kochan}},
  \bibinfo {author} {\bibfnamefont {F.~L.}\ \bibnamefont {Freire~Jr.}}, \ and\
  \bibinfo {author} {\bibfnamefont {W.}~\bibnamefont {Chen}},\ }\href@noop {}
  {\bibinfo  {journal} {arXiv:2405.06146}\ }\BibitemShut {NoStop}%
\bibitem [{\citenamefont {Greenaway}\ and\ \citenamefont
  {Harbeke}(1965)}]{Greenaway65}%
  \BibitemOpen
\bibfield  {journal} {  }\bibfield  {author} {\bibinfo {author} {\bibfnamefont
  {D.}~\bibnamefont {Greenaway}}\ and\ \bibinfo {author} {\bibfnamefont
  {G.}~\bibnamefont {Harbeke}},\ }\href {\doibase
  https://doi.org/10.1016/0022-3697(65)90092-2} {\bibfield  {journal} {\bibinfo
   {journal} {J. Phys. Chem. Solids}\ }\textbf {\bibinfo {volume} {26}},\
  \bibinfo {pages} {1585} (\bibinfo {year} {1965})}\BibitemShut {NoStop}%
\bibitem [{\citenamefont {Chen}(2020)}]{Chen20_absence_edge_current}%
  \BibitemOpen
  \bibfield  {author} {\bibinfo {author} {\bibfnamefont {W.}~\bibnamefont
  {Chen}},\ }\href {\doibase 10.1103/PhysRevB.101.195120} {\bibfield  {journal}
  {\bibinfo  {journal} {Phys. Rev. B}\ }\textbf {\bibinfo {volume} {101}},\
  \bibinfo {pages} {195120} (\bibinfo {year} {2020})}\BibitemShut {NoStop}%
\bibitem [{\citenamefont {Su}\ \emph {et~al.}(1979)\citenamefont {Su},
  \citenamefont {Schrieffer},\ and\ \citenamefont {Heeger}}]{Su79}%
  \BibitemOpen
  \bibfield  {author} {\bibinfo {author} {\bibfnamefont {W.~P.}\ \bibnamefont
  {Su}}, \bibinfo {author} {\bibfnamefont {J.~R.}\ \bibnamefont {Schrieffer}},
  \ and\ \bibinfo {author} {\bibfnamefont {A.~J.}\ \bibnamefont {Heeger}},\
  }\href {\doibase 10.1103/PhysRevLett.42.1698} {\bibfield  {journal} {\bibinfo
   {journal} {Phys. Rev. Lett.}\ }\textbf {\bibinfo {volume} {42}},\ \bibinfo
  {pages} {1698} (\bibinfo {year} {1979})}\BibitemShut {NoStop}%
\bibitem [{\citenamefont {Kitaev}(2001)}]{Kitaev01}%
  \BibitemOpen
  \bibfield  {author} {\bibinfo {author} {\bibfnamefont {A.~Y.}\ \bibnamefont
  {Kitaev}},\ }\href {http://stacks.iop.org/1063-7869/44/i=10S/a=S29}
  {\bibfield  {journal} {\bibinfo  {journal} {Phys. Usp.}\ }\textbf {\bibinfo
  {volume} {44}},\ \bibinfo {pages} {131} (\bibinfo {year} {2001})}\BibitemShut
  {NoStop}%
\bibitem [{\citenamefont {Gradshteyn}\ and\ \citenamefont
  {Ryzhik}(2014)}]{Gradshteyn14}%
  \BibitemOpen
  \bibfield  {author} {\bibinfo {author} {\bibfnamefont {I.~S.}\ \bibnamefont
  {Gradshteyn}}\ and\ \bibinfo {author} {\bibfnamefont {I.~M.}\ \bibnamefont
  {Ryzhik}},\ }\href@noop {} {\emph {\bibinfo {title} {Table of Integrals,
  Series, and Products}}}\ (\bibinfo  {publisher} {Academic Press},\ \bibinfo
  {year} {2014})\BibitemShut {NoStop}%
\bibitem [{\citenamefont {Symon}(1971)}]{Symon71}%
  \BibitemOpen
  \bibfield  {author} {\bibinfo {author} {\bibfnamefont {K.~R.}\ \bibnamefont
  {Symon}},\ }\href@noop {} {\emph {\bibinfo {title} {Mechanics}}}\ (\bibinfo
  {publisher} {Addison-Wesley},\ \bibinfo {year} {1971})\BibitemShut {NoStop}%
\bibitem [{\citenamefont {Bernevig}\ and\ \citenamefont
  {Hughes}(2013)}]{Bernevig13}%
  \BibitemOpen
  \bibfield  {author} {\bibinfo {author} {\bibfnamefont {B.~A.}\ \bibnamefont
  {Bernevig}}\ and\ \bibinfo {author} {\bibfnamefont {T.~L.}\ \bibnamefont
  {Hughes}},\ }\href@noop {} {\emph {\bibinfo {title} {Topological Insulators
  and Topological Superconductors}}}\ (\bibinfo  {publisher} {Princeton
  University Press},\ \bibinfo {year} {2013})\BibitemShut {NoStop}%
\bibitem [{\citenamefont {Bernevig}\ \emph {et~al.}(2006)\citenamefont
  {Bernevig}, \citenamefont {Hughes},\ and\ \citenamefont
  {Zhang}}]{Bernevig06}%
  \BibitemOpen
  \bibfield  {author} {\bibinfo {author} {\bibfnamefont {B.~A.}\ \bibnamefont
  {Bernevig}}, \bibinfo {author} {\bibfnamefont {T.~L.}\ \bibnamefont
  {Hughes}}, \ and\ \bibinfo {author} {\bibfnamefont {S.-C.}\ \bibnamefont
  {Zhang}},\ }\href {\doibase 10.1126/science.1133734} {\bibfield  {journal}
  {\bibinfo  {journal} {Science}\ }\textbf {\bibinfo {volume} {314}},\ \bibinfo
  {pages} {1757} (\bibinfo {year} {2006})}\BibitemShut {NoStop}%
\bibitem [{\citenamefont {Zhang}\ \emph {et~al.}(2009)\citenamefont {Zhang},
  \citenamefont {Liu}, \citenamefont {Qi}, \citenamefont {Dai}, \citenamefont
  {Fang},\ and\ \citenamefont {Zhang}}]{Zhang09}%
  \BibitemOpen
  \bibfield  {author} {\bibinfo {author} {\bibfnamefont {H.}~\bibnamefont
  {Zhang}}, \bibinfo {author} {\bibfnamefont {C.-X.}\ \bibnamefont {Liu}},
  \bibinfo {author} {\bibfnamefont {X.-L.}\ \bibnamefont {Qi}}, \bibinfo
  {author} {\bibfnamefont {X.}~\bibnamefont {Dai}}, \bibinfo {author}
  {\bibfnamefont {Z.}~\bibnamefont {Fang}}, \ and\ \bibinfo {author}
  {\bibfnamefont {S.-C.}\ \bibnamefont {Zhang}},\ }\href {\doibase
  10.1038/nphys1270} {\bibfield  {journal} {\bibinfo  {journal} {Nat. Phys.}\
  }\textbf {\bibinfo {volume} {5}},\ \bibinfo {pages} {438} (\bibinfo {year}
  {2009})}\BibitemShut {NoStop}%
\bibitem [{\citenamefont {Liu}\ \emph {et~al.}(2010)\citenamefont {Liu},
  \citenamefont {Qi}, \citenamefont {Zhang}, \citenamefont {Dai}, \citenamefont
  {Fang},\ and\ \citenamefont {Zhang}}]{Liu10}%
  \BibitemOpen
  \bibfield  {author} {\bibinfo {author} {\bibfnamefont {C.-X.}\ \bibnamefont
  {Liu}}, \bibinfo {author} {\bibfnamefont {X.-L.}\ \bibnamefont {Qi}},
  \bibinfo {author} {\bibfnamefont {H.}~\bibnamefont {Zhang}}, \bibinfo
  {author} {\bibfnamefont {X.}~\bibnamefont {Dai}}, \bibinfo {author}
  {\bibfnamefont {Z.}~\bibnamefont {Fang}}, \ and\ \bibinfo {author}
  {\bibfnamefont {S.-C.}\ \bibnamefont {Zhang}},\ }\href {\doibase
  10.1103/PhysRevB.82.045122} {\bibfield  {journal} {\bibinfo  {journal} {Phys.
  Rev. B}\ }\textbf {\bibinfo {volume} {82}},\ \bibinfo {pages} {045122}
  (\bibinfo {year} {2010})}\BibitemShut {NoStop}%
\bibitem [{\citenamefont {Balian}\ and\ \citenamefont
  {Werthamer}(1963)}]{Balian63}%
  \BibitemOpen
  \bibfield  {author} {\bibinfo {author} {\bibfnamefont {R.}~\bibnamefont
  {Balian}}\ and\ \bibinfo {author} {\bibfnamefont {N.~R.}\ \bibnamefont
  {Werthamer}},\ }\href {\doibase 10.1103/PhysRev.131.1553} {\bibfield
  {journal} {\bibinfo  {journal} {Phys. Rev.}\ }\textbf {\bibinfo {volume}
  {131}},\ \bibinfo {pages} {1553} (\bibinfo {year} {1963})}\BibitemShut
  {NoStop}%
\bibitem [{\citenamefont {Volovik}(2009)}]{Volovik09}%
  \BibitemOpen
  \bibfield  {author} {\bibinfo {author} {\bibfnamefont {G.~E.}\ \bibnamefont
  {Volovik}},\ }\href@noop {} {\emph {\bibinfo {title} {The Universe in a
  Helium Droplet}}}\ (\bibinfo  {publisher} {Oxford University Press},\
  \bibinfo {year} {2009})\BibitemShut {NoStop}%
\end{thebibliography}%

\end{document}